\documentclass[%
 reprint,
 amsmath,amssymb,amsthm,
 nofootinbib,
 aps,
]{revtex4-1}

\usepackage{dcolumn}
\usepackage{bm}
\usepackage[mathlines]{lineno}
\usepackage{graphicx}
\usepackage{csquotes}
\usepackage[final]{changes}

\begin{document}
\preprint{APS/123-QED}

\title{Quantizing gravitational fields with an entropy-corrected action principle}

\author{Jianhao M. Yang}
\email[]{jianhao.yang@alumni.utoronto.ca}
\affiliation{Qualcomm, San Diego, CA 92121, USA}

\date{\today}		

\begin{abstract}
A variational framework for the quantization of gravitational fields is developed based on an extension of the stationary action principle. Within this framework, the Wheeler–DeWitt equation for the gravitational wave functional is recovered without assuming operator promotion of the canonical momentum, thus avoiding the ambiguity of operator ordering in canonical quantization. The derivation is based on three main ingredients. First, motivated by information-theoretic considerations, the classical stationary action principle is generalized by incorporating a correction term constructed from the relative entropy associated with field fluctuations. Second, an ensemble formulation on superspace is enhanced to incorporate this entropy correction. Third, the formalism is further refined to provide a unified treatment of quantization and constraints, thereby addressing the long-standing ambiguity concerning the ordering of quantization and constraint reduction. The framework is then applied to gravitational fields coupled to a massless scalar field. Using an emergent time parameter defined via the rate equation of the gravitational fields, a Schrödinger equation for the scalar-field wave functional is recovered, supplemented by an additional quantum correction term suppressed at order $G\hbar^2$. Finally, we comment on possible connections between the notion of relative entropy employed here and holographic dualities in quantum gravity. 
\end{abstract}

\maketitle
\section{Introduction}
Quantizing gravitational fields within a consistent canonical framework remains one of the central unresolved problems in theoretical physics~\cite{Kiefer, Weinberg}. In the standard ADM formulation of general relativity, the gravitational field is described by a constrained Hamiltonian system on superspace, where the lapse and shift functions enforce the Hamiltonian and momentum constraints associated with spacetime diffeomorphism invariance. Canonical quantization of this system leads to the Wheeler–DeWitt equation, in which the wave functional of the spatial metric is subject to operator-valued constraints rather than a conventional Schrödinger time evolution. This formulation, however, is accompanied by persistent difficulties, including operator-ordering ambiguities, the absence of an external time parameter, and the non-renormalizability of perturbative expansions about fixed backgrounds. In addition, the quantization of constrained Hamiltonian systems introduces further ambiguity, as both reduced phase space quantization~\cite{Ashtekar} and the Dirac quantization procedure involve nontrivial choices regarding the ordering of quantization and constraint implementation~\cite{Dirac, Hanson, Loll, Romano, Schleich, Kunstatter}.

In this work, we develop a framework for quantizing gravitational fields based on an extended stationary action principle augmented by a relative entropy correction. Within this approach, the classical ADM action is generalized by incorporating an information-theoretic term that quantifies fluctuations of field configurations on superspace. The resulting formulation provides a unified variational procedure in which quantization and constraint enforcement are implemented simultaneously. Within this framework, the Wheeler–DeWitt equation is obtained without assuming operator promotion, while the variational structure selects a natural operator ordering and avoids ambiguities associated with the sequencing of quantization and constraint reduction. This provides an alternative route to the quantization of gravitational fields.

The present framework originates from efforts to formulate quantum mechanics on an information-theoretic foundation~\cite{Rovelli:1995fv, zeilinger1999foundational, Brukner:1999qf, Fuchs2002, brukner2009information, Hardy:2001jk, Mueller:2012ai, Masanes:2012uq, chiribella2011informational, Mueller:2012pc, 2008arXiv0805.2770G, Hall2013, Hoehn:2014uua, Hoehn:2015, Caticha2019, Frieden, Reginatto, Yang2021}. This line of research led to the development of an extended stationary action principle~\cite{Yang2023}, which has been shown to reproduce non-relativistic quantum mechanics for both spin-zero and spin-$1/2$ systems~\cite{Yang2023, Yang2025}. Owing to its general formulation within the Lagrangian framework, it is natural to extend this approach to field theory. Previous work has demonstrated that scalar and fermionic fields can be quantized within this framework~\cite{Yang2024, Yang2026}. The objective of the present work is to extend this framework to gravitational fields\footnote{Information-theoretic approaches to quantum field theory and gravity have recently attracted significant attention. For example, gravity models based on entropic actions constructed from quantum relative entropy have been explored in~\cite{Bianconi}.}.

A central element of the extended stationary action principle is the inclusion of an additional term in the classical action, within an ensemble formulation, that accounts for field fluctuations. This term is constructed from a relative entropy functional, which quantifies the information distance between probability distributions with and without fluctuations. By recursively applying the extended stationary action principle, one obtains both the probability density governing these fluctuations and the corresponding Schrödinger equation for the wave functional. In the present work, the framework developed in~\cite{Yang2024, Yang2026} is further extended to incorporate the constraint structure of general relativity directly within the variational principle. In this way, the Hamiltonian constraints of the ADM formulation are treated within the same quantization procedure. We further show that the entropy correction term is invariant under spacetime diffeomorphisms, ensuring that diffeomorphism invariance is preserved at the level of the variational principle.

We further apply the framework to a system of gravitational fields coupled to a massless scalar field. In this setting, an emergent time parameter is defined through the dynamical evolution of the gravitational fields, leading to a Schrödinger equation for the scalar-field wave functional. This equation contains additional correction terms arising from both classical and quantum gravitational effects.

The proposed framework offers both conceptual and mathematical advantages. Conceptually, the entropy-corrected stationary action principle provides a coherent mechanism by which classical field theories are extended to quantum theories through the inclusion of information-theoretic terms that encode field fluctuations. This perspective highlights the role of information measures in the foundations of quantum theory. In particular, recent developments in quantum gravity, including the black hole information paradox~\cite{Mathur} and holographic dualities~\cite{Takayanagi, FLM}, have emphasized the role of entropy. In holographic settings, the relative entropy in a boundary conformal field theory is related to the relative entropy in a bulk gravitational theory~\cite{JLMS}. The incorporation of relative entropy in the present framework may therefore offer new perspectives on these developments. From a mathematical standpoint, the framework is flexible and broadly applicable, as demonstrated in previous work~\cite{Yang2023, Yang2024, Yang2025, Yang2025-2, Yang2026}. It provides an alternative to canonical quantization and path integral formulations. The generality of this framework follows from its basis in the Lagrangian formalism, with the information-theoretic contribution incorporated through the relative entropy functional. The application to gravitational fields presented here further illustrates the scope and adaptability of the approach. 

The mathematical formalism of ensembles in configuration space was originally developed by Hall and Reginatto~\cite{HallBook}. However, the underlying quantization mechanism in this work is via the entropy-corrected action principle, while in ~\cite{HallBook} it is based on an assumption that momentum fluctuations are directly constrained by position uncertainty.

The present work addresses only a subset of the broader challenges in quantum gravity. Issues such as non-renormalizability, the emergence of spacetime, and the black hole information paradox remain beyond the scope of this study.

The remainder of this paper is organized as follows. In Section~\ref{sec:LIP}, we review the underlying assumptions of the extended stationary action principle and formulate the general procedure for quantizing classical fields. Section~\ref{subsec:ADM} reviews the ADM decomposition of the Einstein–Hilbert action, which serves as the starting point for the quantization of gravitational fields. Sections~\ref{sec:shortTimeStep}, \ref{sec:IF}, and \ref{sec:WDE} implement the extended stationary action principle to derive the probability density of field fluctuations and obtain the Wheeler–DeWitt equation. Section~\ref{sec:etime} introduces the emergent time parameter for gravitational fields coupled to a massless scalar field and derives the corresponding Schrödinger-type equation. Finally, Section~\ref{sec:discussion} presents a comparative discussion of quantization approaches and explores potential connections between the relative entropy in this framework and holographic dualities. Technical details are provided in the Appendices.

\section{An Alternative Framework to Quantize a Classical Field}
\label{sec:LIP}
The theoretical framework developed in this paper is based on the extended stationary action principle introduced in~\cite{Yang2023, Yang2024}. In this approach, the classical principle of stationary action is generalized to account for quantum behavior through the incorporation of two additional assumptions:

\begin{displayquote}
\emph{Assumption 1 — Field configurations are subject to intrinsic fluctuations that are random and local.}
\end{displayquote}

\begin{displayquote}
\emph{Assumption 2 — There exists a lower bound on the physically observable action associated with a field configuration. This minimal unit of action is given by $\hbar/2$, where $\hbar$ is the Planck constant.}
\end{displayquote}

The conceptual motivations for these assumptions have been discussed in detail in Refs.~\cite{Yang2023, Yang2024}; here, we briefly summarize the key ideas. The first assumption reflects the standard view that intrinsic fluctuations of field configurations underlie the probabilistic character of quantum phenomena. The presence of such fluctuations necessitates a consistent prescription to evaluate the action functional in a probabilistic setting.

A natural first step is to consider the expectation value of the classical action, obtained by averaging over a probability distribution of field configurations. However, this averaging procedure alone does not fully capture the physical consequences of fluctuations, as it neglects the change in observable information induced by them. Since the stationary action principle determines the dynamical laws governing observable quantities, the effect of fluctuations must be incorporated not only through averaging, but also through their impact on the information content of the system. 

Assumption 2 provides a mechanism for quantifying this additional contribution. It postulates that the effect of unresolved microscopic fluctuations manifests itself through a finite unit of observable action set by $\hbar$. This suggests that the contribution of fluctuations to the total action can be expressed in terms of an information measure characterizing the associated change in observable information. Consequently, we introduce an information functional $I_f$ that quantifies the information content generated by field fluctuations. Multiplication by $\hbar$ then yields the corresponding contribution to the action. In practice, the construction of $I_f$ can be guided by information-theoretic considerations. A natural choice is to define $I_f$ as a functional of a relative entropy measure, such as the Kullback–Leibler divergence $D_{\mathrm{KL}}$, which quantifies the distinguishability between probability distributions associated with fluctuating and non-fluctuating field configurations. We therefore write $I_f := f(D_{\mathrm{KL}})$, where the specific functional form will be specified later.

With these ingredients, the total action functional takes the form
\begin{equation}
\label{totalAction}
    A_t = \langle A_c\rangle + \frac{\hbar}{2}I_f,
\end{equation}
where $\langle A_c \rangle$ denotes the expectation value of the classical action. For notational simplicity, we will hereafter omit the explicit expectation brackets when no ambiguity arises. The dynamics of the system is then obtained by extremizing the total action functional, $\delta A_t = 0$. In the classical limit $\hbar \to 0$, the contribution from $I_f$ vanishes, and the standard stationary action principle is recovered. For $\hbar \neq 0$, the information-theoretic term provides a correction to the classical action and serves as the source of quantum behavior. These considerations lead to the following formulation\footnote{Throughout its development~\cite{Yang2023, Yang2024, Yang2025, Yang2026}, this principle has been described using different terminology, including the principle of least observability and the extended principle of least action, reflecting an evolving understanding of its conceptual basis.}:
\begin{displayquote}
\emph{\textbf{Extended Stationary Action Principle} -- The physical dynamics of a field are determined by the extremization of the total action functional defined in Eq.~(\ref{totalAction}), which includes an information-theoretic correction arising from field fluctuations.}
\end{displayquote}

Appendix~\ref{appdx:qmscalar} provides a detailed implementation of this framework for scalar fields~$\phi$. The general procedure for quantizing a classical field within the extended stationary action framework may be summarized as follows:

\begin{itemize}
    \item \textbf{Step I} Specify the classical theory by writing down the Lagrangian density in the standard form used in canonical field theory.
    \item \textbf{Step II} Apply the extended stationary action principle over an infinitesimal time interval. At this level, the relative entropy is introduced as a local information measure that quantifies the deviation of the probability distribution associated with field fluctuations from a reference distribution (taken here to be uniform). Variation of the total action~\eqref{totalAction} determines the infinitesimal change in the probability density due to fluctuations. This step establishes the local formulation of the information functional and allows one to extract statistical properties of the fluctuations, such as their variance.
    \item \textbf{Step III} Perform a canonical transformation with the Lagrangian density to obtain the Hamilton–Jacobi formulation, characterized by a generating functional $S[\phi, t]$. This requires choosing a foliation of spacetime into spatial hypersurfaces. In Minkowski spacetime, one may take hypersurfaces $\Sigma_t$ of constant time $t$ and introduce a probability density functional $\rho[\phi, t]$ over field configurations on $\Sigma_t$. In curved spacetime, we adopt the ADM decomposition, in which the spatial metric $h_{ij}$ defines the configuration space, and the corresponding probability density functional $\rho[h_{ij}, t]$ is defined on this space. The classical action $A_c$ is then evaluated for the ensemble of configurations.
    \item \textbf{Step IV} Extend the infinitesimal construction of Step II to a finite time interval. The relative entropy is now evaluated between probability densities defined with and without field fluctuations on each hypersurface $\Sigma_t$. Integrating these local contributions over the foliation, for $t \in [0, T]$, yields a global information functional $I_f$. This functional represents the cumulative effect of fluctuations on the action and provides the entropy correction term appearing in Eq.~\eqref{totalAction}.
    \item \textbf{Step V} Perform the variation of the total action to obtain two coupled evolution equations for the functionals $S[\phi, t]$ and $\rho[\phi, t]$. These equations can be combined by introducing the wave functional $\Psi = \sqrt{\rho}e^{iS/\hbar}$, from which a Schrödinger-type equation for $\Psi$ is obtained.
\end{itemize}

Once the Schrödinger equation for the wave functional is obtained and the corresponding Hamiltonian operator is identified, one may recover the standard operator-based formulation of quantum theory. In particular, creation and annihilation operators can be introduced in the usual manner, allowing for the construction of the Fock space and the computation of ground-state and excited-state energies. A key point is that, within the present framework, the Schrödinger equation is not postulated but derived from a variational first principle.

The framework outlined above has already been successfully applied to the quantization of scalar fields~\cite{Yang2024} and fermionic fields~\cite{Yang2026}. In the case of gravitational fields, however, additional difficulties arise due to diffeomorphism invariance in general relativity. As a result, the ADM Hamiltonian is constrained to vanish on-shell and does not generate physical time evolution in the conventional sense. Consequently, the quantization of gravitational fields reduces to the quantization of a constrained Hamiltonian system, where the implementation of constraints plays a central role. 

In the following sections, we extend the framework developed above to address these issues in a consistent manner. In particular, we show that gravitational fields can be quantized within the same extended stationary action principle, with constraints incorporated directly at the level of the variational formulation.


\section{The ADM Decomposition}
\label{subsec:ADM}
The first step in the framework is to write down the proper Lagrangian density for the gravitational fields. We start from the Einstein--Hilbert action (without boundary term)
\begin{equation}
S_{\text{EH}}[g_{\mu\nu}] 
 = \frac{1}{16\pi G} \int_{\mathcal{M}} d^4x \, \sqrt{-g} \, R \,.
\end{equation}
where \(G\) is Newton's gravitational constant, \(g\) is the determinant of the metric tensor \(g_{\mu\nu}\), and \(R\) is the Ricci scalar curvature. Without loss of generality, we omit the cosmological constant $\Lambda$. The ADM $3+1$ decomposition~\cite{ADM1, ADM2} is based on the local foliating spacetime by spacelike hypersurfaces $\Sigma_t$ labeled by a time function $t$. The spacetime metric is written in terms of the lapse function $N(x,t)$, the shift vector $N^i(x,t)$, and the induced 3-metric $h_{ij}(x,t)$ on $\Sigma_t$:
\begin{equation}
\begin{split}
 ds^2 &= g_{\mu\nu} dx^\mu dx^\nu \\
      &= -N^2 dt^2 + h_{ij} (dx^i + N^i dt)(dx^j + N^j dt),
      \end{split}
\end{equation}
where
\begin{equation}
    \begin{split}
        &g^{00} =-\frac{1}{N^2},\text{ }g^{0i} =\frac{N^i}{N^2},\\
        &g^{ij} =h^{ij}-\frac{N^iN^j}{N^2},\\
        &h^{ik}h_{kj}=\delta^i_j,\text{ }\sqrt{-g} =N\sqrt{h}, \text{ } h \equiv\det(h_{ij}).
    \end{split}
\end{equation}
The extrinsic curvature of the hypersurfaces is defined by
\begin{equation}
\label{extCurv}
 K_{ij} = \frac{1}{2N} \left( \dot{h}_{ij} - D_i N_j - D_j N_i \right),
\end{equation}
where $D_i$ is the Levi--Civita covariant derivative associated with 3-metrics $h$. The four-dimensional scalar curvature $R$ can be decomposed as
\begin{equation}
R = {}^{(3)}R + K_{ij} K^{ij} - K^2 + (\text{total boundary term)}.
\end{equation}
where ${}^{(3)}R$ is the scalar curvature of the 3-metrics $h$, and $K = h^{ij} K_{ij}$ is the trace of the extrinsic curvature. The last term is a total divergence and contributes a boundary term to the action. If one adds the appropriate Gibbons--Hawking--York boundary term or neglects boundary contributions, the bulk part of the Einstein--Hilbert action becomes
\begin{equation}
\label{SADM}
S_{\text{ADM}} = \frac{1}{16\pi G} \int dt \, d^3x \, N \sqrt{h} \left( K_{ij} K^{ij} - K^2 + {}^{(3)}R \right).
\end{equation}
One can further simplify the action functional by introducing the DeWitt supermetric, defined as
\begin{equation}
\begin{split}
    G_{ijkl}&=\frac{1}{2}(h_{ik}h_{kl} + h_{il}h_{jk} - h_{ij}h_{kl})\\
    G^{ijkl}&=\frac{1}{2}(h^{ik}h^{kl} + h^{il}h^{jk} - 2h^{ij}h^{kl}).
\end{split}
\end{equation}
The DeWitt supermetrics have the following properties
\begin{equation}
\label{DeWittMetric}
\begin{split}
    &G_{ijkl} =G_{klij},\\
    &G_{ijkl} G^{klmn} =\frac{1}{2}(\delta_i^m\delta^n_j + \delta_i^n\delta^m_j)=\delta^m_{(i}\delta^n_{j)}.    
\end{split}
\end{equation}
With the DeWitt supermetric, the action becomes
\begin{equation}
\label{SADM2}
S_{\text{ADM}} = \frac{1}{16\pi G} \int dt \, d^3x \, N \sqrt{h} \left( K^{ij}G_{ijkl}K^{kl} + {}^{(3)}R \right).
\end{equation}
The canonical momentum conjugate to $h_{ij}$ is
\begin{equation}
\begin{split}
 \pi^{ij} 
   = \frac{\partial \mathcal{L}_{\text{ADM}}}{\partial \dot{h}_{ij}}
   &= \frac{\sqrt{h}}{16\pi G} (K^{ij} - h^{ij} K)\\
   \label{momentumConj}
   &=\frac{\sqrt{h}}{16\pi G} G^{ijkl}K_{kl.}
\end{split}
\end{equation}
In terms of momentum conjugate, the action becomes
\begin{equation}
\label{SADM3}
 S_{\text{ADM}}
   = \int dt \, d^3x (\frac{16\pi GN}{\sqrt{h}} \pi^{ij}G_{ijkl}\pi^{kl}+\frac{N\sqrt{h}}{16\pi G} \, {}^{(3)}R ).
\end{equation}
The total Hamiltonian density is obtained through the Legendre transformation,
\begin{equation}
    \mathcal{H}_{\text{ADM}}=\dot{h}_{ij}\pi^{ij}-\mathcal{L}=\left( N \, \mathcal{H}_{\perp} + N_i \, \mathcal{H}^i \right).
\end{equation}
where the Hamiltonian and momentum constraints are
\begin{align}
\label{constraints}
 \mathcal{H}_{\perp} &= \frac{16\pi G}{\sqrt{h}} \pi^{ij}G_{ijkl}\pi^{kl}
              - \frac{\sqrt{h}}{16\pi G} \, {}^{(3)}R \, , \\
\label{constraintsH_i}
 \mathcal{H}^i &= -2 \, D_j \pi^{ij}.
\end{align}
The lapse $N$ and the shift $N^i$ act as Lagrange multipliers that enforce these constraints. 
The total Hamiltonian is the sum of constraints of the gravitational fields, instead of as the conventional Hamiltonian that generates the time dynamics of the gravitational fields. This is the intrinsic requirement of diffeomorphism invariance in general relativity. The coordinate time is just gauge, not real physical time.

\section{Probability Density of Field Fluctuations}
\label{sec:shortTimeStep}
In Step II, we adopt Assumption 1 that there are intrinsic random fluctuations in the gravitational fields. As a consequence, the gravitation metrics in $\Sigma_t$ fluctuate,
\begin{equation}
    h_{ij}(x,t) \to h_{ij}(x,t) + \Delta h_{ij}(x,t).
\end{equation}
We assume that $\Delta h_{ij}$ is sufficiently small compared to $h_{ij}$ and denote it as $w_{ij}$. To describe the random field fluctuations, we need to define a probability functional $p[h_{ij}+w_{ij}|h_{ij}]$ and a proper measure $\mathcal{D}w$ over the space of metrics $w_{ij}$. We follow the definition in \cite{HallBook} by choosing the DeWitt measure 
\begin{equation}
    \int \mathcal{D}w = \int \prod_x[\sqrt{w(x)}]^{\kappa}\prod_{i\ge j}d w_{ij}(x).
\end{equation}
Without loss of generality, one may set $\kappa=0$ since the term $[\cdot]^{\kappa}$ can be absorbed into the definition of $p$. We require that $p$ meets the normalization condition
$\int \mathcal{D}w p =1$.

With these definitions, the expectation value of the action is
\begin{equation}
    \langle A\rangle_{w} = \int\mathcal{D}w\{ p[h+w|h] S_{\text{ADM}}\},
\end{equation}
where we drop the indices $\{i,j\}$ for $h$ and $w$ in $p[\cdot]$ to simplify the notation. Now we consider an infinitesimal foliation step at $t\to t+\Delta t$. Given Eqs.\eqref{SADM2} and \eqref{extCurv}, when $\Delta t\to 0$, the only surviving terms are
\begin{equation}
    S_{\text{ADM}}=\frac{1}{16\pi G} \int d^3x \frac{\Delta t\sqrt{h}}{4N} ( \dot{h}^{ij}+\dot{w}^{ij})G_{ijkl}(\dot{h}^{kl}+\dot{w}^{kl}).
\end{equation}
For infinitesimal $\Delta t$, we approximate $\dot{w}^{ij}=w^{ij}/\Delta t$, and take the average over $w$,
\begin{equation}
    \langle A\rangle_{w}=\frac{1}{16\pi G} \int  d^3x \frac{\Delta t\sqrt{h}}{4N}\langle( \dot{h}^{ij}+\frac{w^{ij}}{\Delta t})G_{ijkl}(\dot{h}^{kl}+\frac{w^{kl}}{\Delta t})\rangle_w.
\end{equation}
Since the field fluctuation is completely random, it is intuitive to demand that $\langle w^{ij}\rangle = 0$. Thus,
\begin{equation}
    \langle A\rangle_{w}=A_0+\frac{1}{64\pi G} \int \mathcal{D}w \text{ } p\{\int d^3x \frac{\sqrt{h}}{N\Delta t}w^{ij}G_{ijkl}w^{kl}\},
\end{equation}
where $A_0$ is a term independent of $w$ and $p$. We can further express the second term in a compact form of matrix multiplication
\begin{equation}
\label{SADM3}
    \langle A\rangle_{w}=A_0+\int \mathcal{D}w \{p\int d^3x(w\Omega w)\}
\end{equation}
with 
\begin{equation}
\label{Omega}
    \Omega_{ijkl} = \frac{\sqrt{h}}{64\pi GN\Delta t}G_{ijkl}.
\end{equation}

To apply the extended stationary action principle, we need to define the information metric $I_f$. $I_f$ is expected to capture the additional revelation of information due to field fluctuations in $\Sigma_t$. Thus, it is naturally defined as a relative entropy, or more specifically, the Kullback–Leibler divergence, to measure the information distance between $p[w]$ and some prior probability distribution. Given that field fluctuations are completely random, it is intuitive to assume that the prior distribution is with maximal ignorance~\cite{Caticha2019, Jaynes}. That is, the prior probability distribution is a uniform distribution $\sigma$. 
\begin{align*}
    I_f  &:= D_{KL}(p[h+w|h]|| \sigma) \\
    &= \int \mathcal{D}w p[h+w|h]ln(\frac{p[h+w|h]}{\sigma}).
\end{align*}
Combined with \eqref{SADM3}, the total action functional defined in (\ref{totalAction}) is given by
\begin{align*}
    A_t = A_0+\int \mathcal{D}w \{p\int d^3x(w\Omega w)+\frac{\hbar}{2}p\ln\frac{p}{\sigma}\}.
\end{align*}
Adding the constraint due to normalization of $p$, and performing variation $\delta A_t=0$ with respect to $p$, we find that
\begin{equation}
\label{transProb}
    p[h+w|h] = \frac{1}{Z}\exp\{-\frac{2}{\hbar}\int d^3x(w\Omega w)\},
\end{equation}
where $Z$ is the normalization factor. We recognize that $p$ is independent of $h$ and therefore $p$ can be denoted as $p[w]$. Eq.\eqref{transProb} can be written in the well-known form of Gaussian functional, 
\begin{equation}
\label{transProb2}
    p[w] = \frac{1}{Z}\exp\{-\frac{1}{2}\int d^3x(w\bar\Omega w)\},
\end{equation}
where $\bar\Omega=(4/\hbar)\Omega$ is the quadratic kernel. For a Gaussian functional with a quadratic kernel $\bar\Omega$, the following identities hold~\cite{ZeeBook}
\begin{equation}
    \langle w^{ij}(x)\rangle = 0, \text{ } \langle w^{ij}(x)w^{kl}(y)\rangle=\delta(x-y)\bar\Omega^{-1}_{ijkl}.
\end{equation}
Substituting \eqref{Omega} into $\bar\Omega$, we have
\begin{equation}
    \langle w^{ij}(x)w^{kl}(y)\rangle=\frac{16\pi GN\hbar\Delta t}{\sqrt{h}}G^{ijkl}\delta(x-y).
\end{equation}
Similarly,
\begin{equation}
\label{varianceW}
    \langle w_{ij}(x)w_{kl}(y)\rangle=\frac{16\pi GN\hbar\Delta t}{\sqrt{h}}G_{ijkl}\delta(x-y).
\end{equation}
From \eqref{momentumConj}, one can compute the fluctuation of momentum conjugate due to field fluctuation in an infinitesimal slice of time as, up to $O(\Delta h)$,
\begin{equation}
\begin{split}
     \Delta\pi^{mn} &= \frac{\sqrt{h}}{16\pi G}G^{mnkl}\Delta K_{kl}=\frac{\sqrt{h}}{32\pi GN}G^{mnkl}\Delta\dot{h}_{kl}\\ &=\frac{\sqrt{h}}{32\pi GN}G^{mnkl} \frac{w_{kl}}{\Delta t}.
\end{split}
\end{equation}
Substituting it into \eqref{varianceW}, and using the properties of \eqref{DeWittMetric}, we find that
\begin{equation}
     \langle \Delta h_{ij}(x)\Delta\pi^{mn}(y)\rangle=\frac{\hbar}{2}\delta^m_{(i}\delta^n_{j)}\delta(x-y).
\end{equation}
Applying the Cauchy-Schwarz inequality, we have
\begin{equation}
\begin{split}
     \langle \Delta h_{ij}(x)\rangle\langle\Delta\pi^{mn}(y)\rangle &\ge \langle \Delta h_{ij}(x)\Delta\pi^{mn}(y)\rangle\\&=\frac{\hbar}{2}\delta^m_{(i}\delta^n_{j)}\delta(x-y).
\end{split}
\end{equation}
However, comparing the LHS to the delta function in the RHS appears mathematically inappropriate. Instead, we introduce a pair of positive spatial test functions $f(x), g(y):\mathbb{R}^3\to\mathbb{R}^+$, and define
\begin{equation}
\begin{split}
    &\langle w^{ij}(f)w^{kl}(g)\rangle:= \\&\int\mathcal{D}wp[w]\int d^3xd^3y w^{ij}(x)f(x)w^{kl}(y)g(y).
\end{split}
\end{equation}
Repeating the calculation, we obtain
\begin{equation}
     \langle \Delta h_{ij}(f)\rangle\langle\Delta\pi^{mn}(g)\rangle \ge \frac{\hbar}{2}\delta^m_{(i}\delta^n_{j)}\langle f|g\rangle.
\end{equation}
where $\langle f|g\rangle=\int d^3xf(x)g(x)$. This is the uncertainty relation between the gravitational field variable $h$ and its momentum conjugate $\pi$, due to random fluctuations of the gravitational field. Eq.\eqref{varianceW} plays a crucial role in the quantization calculation later. 

\section{Ensemble of Classical Gravitational Fields}
\label{sec:extCanTrans}
In Step III, we apply the canonical transformation technique in field theory to transform the action functional so that the Hamilton-Jacobi equation can be obtained. The canonical transformation process is similar to that for scalar fields as shown in Appendix \ref{appdx:qmscalar}. The important step is to introduce a generating functional $S[h_{ij}, t]$ in gravitational field configurations that satisfies
\begin{align}
\label{Conjugate}
   \pi^{ij} = \frac{\delta S}{\delta h_{ij}}.
\end{align}
Note that $\pi^{ij}$ is a tensor density of weight +1 therefore $\delta S/\delta h_{ij}$ is also a tensor density of weight +1. Substituting \eqref{Conjugate} into \eqref{constraints} and \eqref{constraintsH_i}, we obtain
\begin{align}
 \mathcal{H}_{\perp} &= \frac{16\pi G}{\sqrt{h}} \frac{\delta S}{\delta h_{ij}}G_{ijkl}\frac{\delta S}{\delta h_{kl}}
              - \frac{\sqrt{h}}{16\pi G} \, {}^{(3)}R \, , \\
 \mathcal{H}^i &= -2 D_j (\frac{\delta S}{\delta h_{ij}}).
\end{align}

The action functional, after the canonical transformation, becomes
\begin{equation}
\label{action10}
\begin{split}
    A'_{\text{ADM}}&=\int dt\{\frac{\partial S}{\partial t}+\int d^3x\mathcal{H}_{\text{ADM}}\}\\
    &=\int dt\{\frac{\partial S}{\partial t}+\int d^3x(N\mathcal{H}_{\perp}+N_i\mathcal{H}^i)\}.
\end{split}
\end{equation}
Suppose a particular type of $S$ is chosen to satisfy the Einstein-Hamilton-Jacobi(EHJ) equation 
\begin{equation}
    \frac{\partial S}{\partial t}+\int d^3x\mathcal{H}_{\text{ADM}} = 0,
\end{equation}
then $\delta A'_{\text{ADM}}=0$ automatically. Therefore, the EHJ equation can be considered as a special solution when applying the variation procedure on the action functional $A'_{\text{ADM}}$. 
Performing variations of $A'_{\text{ADM}}$ with respect to $N$ and $N_i$ gives
\begin{align}
    \mathcal{H}_{\perp} \approx 0; \text{ }\mathcal{H}^i \approx 0.
\end{align}
From the EHJ equation it immediately follows that 
\begin{equation}
\label{STcontraint}
    \frac{\partial S}{\partial t} \approx 0. 
\end{equation}
Alternatively, constraints $\partial S/\partial t \approx0 $ and $\mathcal{H}^i\approx 0$ can be obtained by imposing the requirement that the generating functional $S$ must be invariant under time coordinate reparameterization or under spatial coordinate transformation. The proof of $\mathcal{H}^i\approx 0$ from the requirement of $S$ is invariant under spatial transformation is provided in Appendix \ref{appdx:gravityconstraint}. 

Fluctuations in the gravitational field introduce uncertainty for $h_{ij}$. To deal with uncertainty, we consider an ensemble of field configurations $h_{ij}$ in a hypersurface $\Sigma_t$ at each spatial point. Define the DeWitt measure on the configuration space of $h$ as (with $\kappa=0$)
\begin{equation}
    \int \mathcal{D}h = \int \prod_x[\sqrt{h(x)}]^{\kappa}\prod_{i\ge j}d h_{ij}(x),
\end{equation}
and a probability distribution with probability density $\rho[h,t]$. Then, given \eqref{action10}, the expectation value of the action functional for the ensemble is
\begin{equation}
    \label{ensembleAction}
    \langle A_c\rangle = \int dt\mathcal{D}h \{\rho (\frac{\partial S}{\partial t} + \int d^3x\mathcal{H}_{\text{ADM}})\}.
\end{equation}
The pair of functionals $(\rho, S)$ can be treated as generalized canonical variables~\cite{Hall:2001, Yang2023, Yang2024, Yang2025}. In the super-space defined by $(\rho, S)$, the Lagrangian density and Hamiltonian of the ensemble is
\begin{align}
    \mathcal{L}_{\text{en}} &= \rho (\frac{\partial S}{\partial t} + \int d^3x\mathcal{H}_{\text{ADM}}),\\
    \label{enH}
    \mathcal{H}_{\text{en}} &= \rho\dot{S}-\mathcal{L}_{\text{en}} =-\rho\int d^3x\mathcal{H}_{\text{ADM}}.
\end{align}

Similarly to the constraints we impose on $S$, we impose the same constraint on $\rho$, that $\rho$ must be invariant under time coordinate reparameterization or under spatial coordinate transformation. This requirement leads to 
\begin{align}
\label{Pconstraints}
    \frac{\partial \rho}{\partial t} \approx 0,\text{ }
    D_j (\frac{\delta \rho}{\delta h_{ij}}) &\approx 0.
\end{align}
In summary, we have identified four sets of constraints. There are two constraints for $\rho$ as specified in \eqref{Pconstraints}, and the other two for $S$, repeated here for clarity,
\begin{align}
\label{Sconstraints}
    \frac{\partial S}{\partial t} \approx 0,\text{ }
    D_j (\frac{\delta S}{\delta h_{ij}}) &\approx 0.
\end{align}
In Appendix \ref{appdx:gravityconstraint}, we confirm that these constraints do not induce other secondary constraints.

We can apply the stationary action principle to the action functional defined in (\ref{ensembleAction}). The variation of $\langle A_c\rangle$ with respect to $\rho$ leads to the EHJ equation. Taking into account the constraints in \eqref{Sconstraints}, the EHJ equation becomes
\begin{equation}
    \label{EHJ2}
    \frac{16\pi G}{\sqrt{h}} \frac{\delta S}{\delta h_{ij}}G_{ijkl}\frac{\delta S}{\delta h_{kl}}
              - \frac{\sqrt{h}}{16\pi G} \, {}^{(3)}R  =0.
\end{equation}
The variation of $\langle A_c\rangle$ with respect to $S$ gives an equation equivalent to the continuity equation for the probability density $\rho$. Taking into account the constraints in \eqref{Pconstraints}, the continuity equation becomes
\begin{equation}
    \label{Cont2}
    \int d^3x\frac{N}{\sqrt{h}} \frac{\delta }{\delta h_{ij}}\{\rho G_{ijkl}\frac{\delta S}{\delta h_{kl}}\}=0.
\end{equation}
as shown later in Appendix \ref{appendix:SE}. The EHJ equation and the continuity equation together determine the dynamics of the ensemble of classical gravitational fields under constraints \eqref{Pconstraints} and \eqref{Sconstraints}. 

\section{Quantization of the Field Ensembles}
\subsection{Relative Entropy of the Field Ensembles}
\label{sec:IF}
To apply the extended stationary action principle, first we compute the action functional from the dynamics of the classical field ensemble with probability density $\rho[h,t]$ as defined in (\ref{ensembleAction}). Next, in Step IV,  we define the information metrics for the field fluctuations, $I_f$. Consider the field dynamics for a period of time from $t_A\to t_B$. The spacetime from $t_A\to t_B$ is sliced into a succession of $N$ local Cauchy hypersurfaces $\Sigma_{t_i}$, where $t_i \in \{t_0=t_A, \ldots, t_i, \ldots, t_{N-1}=t_B\}$, and each time step is an infinitesimal period $\Delta t$. The field configuration for each $\Sigma_{t_i}$ is denoted as $h(t_i)$, which has an infinite number of components, labeled as $h_{ij,x}(t_i)=h_{ij}(x, t_i)$, for each spatial point in $\Sigma_{t_i}$. For each field configuration ($h+w$) where $w$ is the change in field configuration due to field fluctuations between $t_i$ and $t_i+\Delta t$, there is a different probability density $\rho[h+w, t_i]$. Consequently, there is additional revelation of information due to the field fluctuations in addition to the dynamics of the classical field ensemble. The proper measure of this distinction is the information distance between $\rho[h, t_i]$ and $\rho[h+w, t_i]$. A natural choice for such an information measure is the relative entropy $D_{KL}(\rho[h, t_i] || \rho[h+w, t_i])$. We need to consider the contributions for all possible $w$, therefore taking the expectation value of $D_{KL}$ over $w$, denoted as $\langle\cdot\rangle_{w}$. Thus, the contribution of additional information due to field fluctuations in the hypersurface $\Sigma_{t_i}$ is $\langle D_{\text{KL}}(\rho[h, t_i] || \rho[h+w, t_i])\rangle_{\omega}$. Finally, we sum up the contributions from all the hypersurfaces, and obtain the definition of information metrics
\begin{align}
\label{DLDivergence2}
    I_f &:= \sum_{i=0}^{N-1}\langle D_{\text{KL}}(\rho[h, t_i] || \rho[h+h, t_i])\rangle_{w} \\
    &=\sum_{i=0}^{N-1}\langle\int \mathcal{D}h \{\rho [h, t_i]\ln \frac{\rho[h, t_i]}{\rho [h+w,t_i]}\}\rangle_w.
\end{align}
As shown in Appendix \ref{appendix:I_f}, with the help of \eqref{varianceW}, when $\Delta t\to 0$, $I_f$ turns out to be
\begin{equation}
\label{FisherInfo2}
    I_f = \frac{\hbar}{2}\int dtd^3x\mathcal{D}h \{\frac{16\pi GN}{\sqrt{h}}\frac{1}{\rho}\frac{\delta\rho}{\delta h_{ij}}G_{ijkl}\frac{\delta\rho}{\delta h_{kl}} \} .
\end{equation}
Eq. (\ref{FisherInfo2}) is analogous to the Fisher information for the probability density~\cite{Yang2023, Frieden}. Some previous literature \cite{Frieden, Reginatto} introduced Fisher information directly in the quantization process. Although the connection between Fisher information and the relative entropy is quite obvious, to prove the exact relation between the two quantities, one needs the identity in \eqref{varianceW}, which is non-trivial. What is elegant in our framework is that the identity \eqref{varianceW} itself can be derived through the same extended principle of least action, as shown in Section \ref{sec:shortTimeStep}.

Two comments are worth mentioning here. First, defining $I_f$ using relative entropy opens up new results that cannot be obtained if $I_f$ is defined using Fisher information, because there are other generic forms of relative entropy such as R\'{e}nyi divergence or Tsallis divergence. Second, given $\rho$ is diffeomorphism invariant, it follows that $I_f$ is also diffeomorphism invariant.

\subsection{Quantization with Constraints}
\label{subsec:qconstraint}
The constraints \eqref{Pconstraints} and \eqref{Sconstraints} pose additional challenges when we quantize the gravitational fields. The central issue is the ordering of solving constraints versus quantization. The \textit{reduced quantization}~\cite{Isham} solves the constraints first at the classical level, then quantizes the reduced system. The \textit{Dirac quantization}~\cite{Dirac}, on the other hand, quantizes the unconstrained system first, then the constraints are solved at the quantum level. Whether quantization and reduction commute is a subtle and highly nontrivial question, and in general they do not. Numerous studies have shown explicit cases where the two approaches yield inequivalent quantum theories\cite{Ashtekar, Loll, Romano, Schleich, Kunstatter}. Quantizing the constraints separately can introduce quantum effects that are absent in the classically reduced theory. However, no rigorous proof exists that constraint quantization is universally self-consistent. 

Traditionally, the Wheeler-DeWitt equation is derived through the reduced quantization approach. Constraints are enforced at the classical level, and the reduced classical Hamiltonians are obtained as $\mathcal{H}_{\perp}$ and $\mathcal{H}^i$. Then, standard canonical quantization is applied by promoting the momentum conjugate as an operator $\pi^{ij} \to -i\hbar\frac{\delta}{\delta h_{ij}}$.
Substituting it into $\mathcal{H}_{\perp}$ and $\mathcal{H}^i$, one can obtain the Wheeler-DeWitt equation.

The mathematical framework presented in Section II does not have this mathematical step of promoting the momentum to be an operator. Instead, it relies on the variation method to derive the Schr\"{o}dinger equation of the wave functional for the fields. This framework allows us to naturally integrate the quantization and the constraint enforcement simultaneously. We quantize the Hamiltonian and the constraints in the same steps when we perform the variation procedure using the extended stationary action principle. The treatment of constraints for a classical ensemble is explained in Appendix \ref{appdx:qconst}. Here, we show how the framework is extended to a quantum ensemble by including $I_f$. For convenience, we define a functional $F[\rho]$ from
\begin{equation}
    \int dt \mathcal{D}h \text{ } \rho F[\rho] = \frac{\hbar}{2}I_f,
\end{equation}
where $I_f$ is given by \eqref{FisherInfo2}. Adding the term $I_f$ to the action functional in \eqref{ensembleAction}, the classical ensemble becomes a quantum ensemble. If there are no constraints, the Lagrangian density of the quantum ensemble is
\begin{equation}
    \mathcal{L}_{\text{q}}=\rho(\dot{S}+H[h,\frac{\delta S}{\delta h}]+F[\rho]).
\end{equation}
We then perform the variation on the action functional with respect to $S$ and $\rho$ to derive the Schr\"{o}dinger equation.

Now consider a set of constraints, $\mathcal{C}_\alpha$, $\alpha=1,2,3,\ldots$, such as those identified in Appendix \ref{appdx:gravityconstraint}. We simply add these constraints to the Lagrangian density with the Langrage multipliers $\lambda^{(\alpha)}$. Define $\Phi_\alpha[\rho, S]$ such that
\begin{equation}
    \mathcal{C}_\alpha =\int \mathcal{D}h \text{ }\rho[h]\Phi_\alpha[\rho, S],
\end{equation}
the Lagrangian density of the quantum ensemble with constraint becomes
\begin{equation}
    \mathcal{L}_{\text{qc}}=\rho(\dot{S}+H[h,\frac{\delta S}{\delta h}] +F[\rho]+\lambda^{(\alpha)}\Phi_a).
\end{equation}
Performing the variation on the action functional with respect to $S$, $\rho$, and $\lambda^\alpha$, allows us to derive the constrained version of the Schr\"{o}dinger equation. It is possible that the constrained Schr\"{o}dinger equation derived from this procedure is the same as that from the Dirac quantization or reduced quantization approaches. It depends on the nature of the constraints. Example of different outcomes from the present quantization framework and the Dirac or reduced quantization can be found in~\cite{Yang2025-2}. 

\subsection{The Wheeler-DeWitt Equation} 
\label{sec:WDE}
We now have all the elements to write down the total action functional that includes the relative entropy term and the constraints identified in Appendix \ref{appdx:gravityconstraint}.
\begin{align*}
    A_t =& \int dt\mathcal{D}h\text{ }\mathcal{L}_\text{qc}\\
    =&\int dt\mathcal{D}h\text{ }\rho\{\frac{\partial S}{\partial t}+\int d^3x(N\mathcal{H}_\perp+N_i\mathcal{H}^i) \\
    &+F[\rho]+\lambda^{(1)}\frac{\partial S}{\partial t}+\lambda^{(3)}\frac{\partial \rho}{\partial t}\\
    &+\int d^3x[\lambda^{(2)}_iD_j(\frac{\delta S}{\delta h_{ij}})+\lambda^{(4)}_iD_j(\frac{\delta \rho}{\delta h_{ij}})]\}.
\end{align*}
Performing the variation on $A_t$ with respect to $N(x,t)$, $N_i(x,t)$, and $\lambda^\alpha(x,t)$, respectively, we obtain the following equations, listed side-by-side with the variation being performed:
\begin{align}
\label{var1}
    &\delta_{\lambda^{(1)}}A_t:\text{ }\frac{\partial S}{\partial t} = 0,\\
\label{var2}
    &\delta_{\lambda^{(2)}_i}A_t:\text{ } D_j(\frac{\delta S}{\delta h_{ij}})=0,\\
\label{var3}
    &\delta_{\lambda^{(3)}}A_t:\text{ }\frac{\partial \rho}{\partial t} = 0,\\
\label{var4}
    &\delta_{\lambda^{(4)}_i}A_t:\text{ } D_j(\frac{\delta \rho}{\delta h_{ij}})=0,\\
\label{var5}
    &\delta_{N_i}A_t:\text{ } \mathcal{H}^i = -2\sqrt{h} D_j(\frac{\delta S}{\delta h_{ij}})=0,\\
\label{var6}
    &\delta_{N}A_t:\text{}\int d^3x\mathcal{D}h\text{}(\rho\mathcal{H}_\perp+\frac{4\pi G\hbar^2}{\sqrt{h}}\frac{1}{\rho}\frac{\delta\rho}{\delta h_{ij}}G_{ijkl}\frac{\delta\rho}{\delta h_{kl}})=0.
\end{align}
Variations of $A_t$ with respect to $S$ and $\rho$ are more complicated. We show the detailed calculation in Appendix \ref{appendix:SE}, and summarize the result here. Variation with respect to $S$ results in
\begin{align}
\label{var7}
&(1+\lambda^{(1)})\frac{\partial\rho}{\partial t}+32\pi G\int d^3xN\frac{\delta }{\delta h_{ij}}\{\rho \frac{G_{ijkl}}{\sqrt{h}} \frac{\delta S}{\delta h_{kl}}\}=0,
\end{align}
while variation with respect to $\rho$ gives
\begin{equation}
\label{var8}
\begin{split}
    &(1+\lambda^{(1)})\frac{\partial S}{\partial t}+\int d^3x[N(\mathcal{H}_\perp+Q)+(N_i+\lambda_i^{(2)})\mathcal{H}^i]=0,
\end{split}
\end{equation}
where
\begin{equation}
    \label{BohmPotential}
    Q[\rho]=-\frac{16\pi G\hbar^2}{\sqrt{\rho}}\frac{\delta }{\delta h_{ij}}(\frac{G_{ijkl}}{\sqrt{h}}\frac{\delta \sqrt{\rho}}{\delta h_{kl}})\}.
\end{equation}
$Q[\rho]$ can be interpreted as the Bohm quantum potential~\cite{Bohm1952, Bohm2} for gravitational fields due to random field fluctuations.

Eqs.\eqref{var1}-\eqref{var8} are the complete results of extremizing the total action functional with constraints. The effects of constraints are intrinsically built into the variation process. In this sense, we quantize the Hamiltonian and the constraints in the same time. The remaining task is just to solve these equations. 

Eqs.\eqref{var1}-\eqref{var4} simply recover the constraints themselves. Eqs.\eqref{var2} and \eqref{var5} are the same. Substitute \eqref{var1}, \eqref{var3} and \eqref{var5} into \eqref{var7} and \eqref{var8}, \eqref{var7} and \eqref{var8} are simplified as
\begin{align}
\label{var9}
    &\int d^3xN \frac{\delta }{\delta h_{ij}}(\rho \frac{G_{ijkl}}{\sqrt{h}}\frac{\delta S}{\delta h_{kl}})=0,\\
\label{var10}
    &\int d^3xN(\mathcal{H}_\perp +Q) = 0. 
\end{align}
Performing variation of \eqref{var6} with respect to $\rho$, one obtains \eqref{var10}. Therefore, \eqref{var6} and \eqref{var10} are equivalent. Since $N(x,t)$ is an arbitrary multiplier, the integrands in \eqref{var9}-\eqref{var10} must be set to zero. Thus,
\begin{align}
\label{var11}
    &\frac{\delta }{\delta h_{ij}}(\rho \frac{G_{ijkl}}{\sqrt{h}}\frac{\delta S}{\delta h_{kl}})=0,\\
\label{var12}
    &\mathcal{H}_\perp+Q=0.
\end{align}

In Summary, there are only six independent equations, \eqref{var1}-\eqref{var4}, and \eqref{var11}-\eqref{var12} after the quantization procedure. Define a complex functional 
\begin{equation}
\label{wavefunctional}
    \Psi[h,t]=\sqrt{\rho[h,t]}\exp{(\frac{i}{\hbar}S[h, t])}.
\end{equation}
Eqs. \eqref{var1} and \eqref{var3} are equivalent to
\begin{equation}
\label{Psi1}
    \frac{\partial \Psi}{\partial t} = 0,
\end{equation}
and Eqs. \eqref{var2} and \eqref{var4} give
\begin{equation}
\label{Psi2}
    D_j(\frac{\delta \Psi}{\delta h_{ij}}) = 0.
\end{equation}
These two equations simply reaffirm that the wave functional $\Psi$ is invariant under spacetime differomorphism. Finally, Eqs. \eqref{var11} and \eqref{var12} are combined into
\begin{equation}
\label{Psi3}
    \{-16\pi G\hbar^2\frac{\delta }{\delta h_{ij}}(\frac{G_{ijkl}}{\sqrt{h}}\frac{\delta }{\delta h_{kl}})
              - \frac{\sqrt{h}}{16\pi G} \, {}^{(3)}R\}\Psi  =0.
\end{equation}
Eq.\eqref{Psi3} is almost the same as the Wheeler-DeWitt equation for the wave functional of gravitational fields derived from canonical quantization, with a difference that the factor $\frac{G_{ijkl}}{\sqrt{h}}$ must be rigorously placed between the two functional derivative operators $\frac{\delta }{\delta h_{ij}}$. The difference is important because the factor $\frac{G_{ijkl}}{\sqrt{h}}$ is a functional of the field variable $h$ and its functional derivative generally does not vanish. This form of the Wheeler-DeWitt equation avoids the operator ordering ambiguity compared to the Wheeler-DeWitt equation obtained from canonical quantization, as pointed out by Hall and Reginatto.  

The three equations \eqref{Psi1}-\eqref{Psi3} give a complete description of the quantum behavior of gravitational fields. These results are identical to those obtained using the approach of canonical quantization of the constrained system. The reason for this is that the constraints \eqref{var1}-\eqref{var4} can be expressed as linear operators acting on the wave functional, as shown in \eqref{Psi1}-\eqref{Psi2}. When constraints cannot be expressed as linear operators acting on the wave functional, the results of the quantization of our framework and the canonical quantization approach can be different~\cite{Yang2025-2}.

\subsection{Generalized Relative Entropy}
\label{sec:Tsallis}
The term $I_f$ is supposed to capture additional information exhibited by field fluctuations and is defined in (\ref{DLDivergence2}) as the summation of the expectation values of the Kullback-Leibler divergence between $\rho[h,t]$ and $\rho[h+w, t]$. However, there are more general definitions of relative entropy, such as the Tsallis divergence~\cite{Tsallis, Nielsen2011}. From an information-theoretic point of view, it is legitimate to consider alternative definitions of relative entropy. Suppose that we define $I_f$ based on Tsallis divergence\footnote{It is possible to choose another one-parameter generalization of relative entropy, such as R\`{e}nyi divergence~\cite{Renyi, Erven2014}. We expect that the result will be similar, as in the case of quantization of the scalar field \cite{Yang2024}.}
\begin{align}
    &I_f^\alpha := \sum_{i=0}^{N-1}\langle D_T^\alpha(\rho[h, t_i] || \rho[h+w, t_i])\rangle_{\omega} \\
    \label{TDivergence}
    &=\sum_{i=0}^{N-1}\langle\frac{1}{\alpha-1}(\int \mathcal{D}\psi^\dagger\mathcal{D}\psi \frac{\rho^\alpha[h, t_i]}{\rho^{\alpha-1} [h+w,t_i]} - 1)\rangle_\omega.
\end{align}
The parameter $\alpha \in (0,1)\cup(1, \infty)$ is called the order of Tsallis divergence. In Appendix \ref{appendix:I_f}, we show that when $\Delta t\to 0$,
\begin{equation}
\label{Tsallis}
    I_f^\alpha = \frac{\alpha\hbar}{2} \int\mathcal{D}h d^3xdt\{\frac{16\pi GN}{\rho}\frac{\delta\rho}{\delta h_{ij}}\frac{G_{ijkl}}{\sqrt{h}}\frac{\delta\rho}{\delta h_{kl}}\} = \alpha I_f.
\end{equation}
When $\alpha\to 1$, $I_f^{\alpha}$ converges to $I_f$ as defined in (\ref{FisherInfo}), as expected.

The parameter $\alpha$ provides a new degree of freedom when we quantize gravitational fields. Using \eqref{Tsallis}, and following the same calculation steps as in Appendix \ref{appendix:SE}, we find that the quantum Einstein-Hamilton-Jacobi equation becomes
\begin{equation}
\label{QHJ2}
    \mathcal{H}_\perp-16\pi G\hbar^2\frac{\alpha}{\sqrt{\rho}} \frac{\delta }{\delta h_{ij}}( \frac{G_{ijkl}}{\sqrt{h}}\frac{\delta \sqrt{\rho}}{\delta h_{kl}})=0.
\end{equation}
The continuity equation \eqref{var11} stays the same. We now define the wave functional as
\begin{equation}
\label{wavefunctional2}
    \Psi[h,t]=\sqrt{\rho[h,t]}\exp{(\frac{i}{\sqrt{\alpha}\hbar}S[h, t])},
\end{equation}
the Wheeler-DeWitt equation becomes
\begin{equation}
\label{Psi4}
    \{-16\alpha\pi G\hbar^2\frac{\delta }{\delta h_{ij}}(\frac{G_{ijkl}}{\sqrt{h}}\frac{\delta }{\delta h_{kl}})
              - \frac{\sqrt{h}}{16\pi G} \, {}^{(3)}R\}\Psi  =0.
\end{equation}
It is not clear at this point whether there is physical significance to this generalization of the Wheeler-DeWitt equation. Eq.\eqref{Psi4} just shows the mathematical flexibility of using the relative entropy to define the information metrics $I_f$, instead of using Fisher information.

\section{The Emergent Time}
\label{sec:etime}
The Wheeler-DeWitt equation and Eq.\eqref{Psi1} show that the wave functional does not evolve with time. They are instead constraints imposed on the wave functional. This is the well-known ``problem of time"~\cite{Anderson}. It is a consequence of the fundamental difference between standard quantum mechanics, where time is absolute, and general relativity, where time is dynamical and relative. One solution is to define physical ``time" from certain internal degrees of freedom or semiclassical approximations. Here, we want to investigate whether the quantization framework presented here can address the problem of time. A non-perturbative example to illustrate how the time emerges is to study the gravitational fields coupled with massless scalar fields~\cite{HallBook}. The idea is to read off the time parameter from the rate equation of the gravitational fields and treat it as a relative clock, which in turn allows us to derive the Schr\"{o}dinger type of dynamical equation for the wave functional of the scalar fields. 


We start the quantization process by writing down the classical Lagrangian density of the combined gravitational fields and massless scalar fields.
\begin{equation}
    \mathcal{L} = \frac{1}{16\pi G}\int d^4x \sqrt{-g}R + \frac{1}{2}\int d^4x \sqrt{-g}g^{\mu\nu}\partial_\mu\phi\partial_\nu\phi.
\end{equation}
After the ADM 3+1 decomposition, the Lagrangian density becomes
\begin{equation}
\begin{split}
    \mathcal{L}_\text{ADM}=&\frac{N\sqrt{h}}{16\pi G}(K_{ij}K^{ij}-K^2+R^{(3)})\\
    &+\frac{\sqrt{h}}{2N}(\dot{\phi}-N^i\partial_i\phi)^2-\frac{\sqrt{h}}{2}h^{ij}\partial_i\phi\partial_j\phi.
\end{split}
\end{equation}
Using the momentum conjugates defines as
\begin{align}
    \pi^{ij} &=\frac{\sqrt{h}}{16\pi G}G^{ijkl}K_{kl},\\
    p_\phi &=\frac{\sqrt{h}}{N}(\dot{\phi}-N^i\partial_i\phi),
\end{align}
we find that the total Hamiltonian density becomes
\begin{align}
\mathcal{H}_\text{tot}&=N(\mathcal{H}_g+\mathcal{H}_\phi)+N_i\mathcal{H}^i,\\
    \mathcal{H}_g &=\frac{16\pi G}{\sqrt{h}}\pi^{ij}G_{ijkl}\pi^{kl} -\frac{\sqrt{h}}{16\pi G}R^{(3)},\\
    \mathcal{H}_\phi&=\frac{1}{2\sqrt{h}}p^2_\phi+\frac{\sqrt{h}}{2}h^{ij}\partial_i\phi\partial_j\phi,\\
     \mathcal{H}^i &=-D_j\pi^{ij} +p_\phi\partial^i\phi.
\end{align}
Introduced the generating functional $S(h,\phi)$ that satisfies
\begin{equation}
\label{momentumConj2}
    \pi^{ij} = \frac{\delta S}{\delta h_{ij}}, \text{ } p_\phi=\frac{\delta S}{\delta \phi},
\end{equation}
the Hamiltonian densities are rewritten as
\begin{align}
    \mathcal{H}_g[S] &=\frac{16\pi G}{\sqrt{h}}\frac{\delta S}{\delta h_{ij}}G_{ijkl}\frac{\delta S}{\delta h_{kl}} -\frac{\sqrt{h}}{16\pi G}R^{(3)},\\
    \mathcal{H}_\phi [S]&=\frac{1}{2\sqrt{h}}(\frac{\delta S}{\delta \phi})^2+\frac{\sqrt{h}}{2}h^{ij}\partial_i\phi\partial_j\phi,\\
     \mathcal{H}^i[S] &=-D_j(\frac{\delta S}{\delta h_{ij}}) +\frac{\delta S}{\delta \phi}\partial^i\phi.
\end{align}
The Einstein-Hamilton-Jacobi equation retains the well-known form,
\begin{equation}
    \frac{\partial S}{\partial t}+\int d^3 \mathcal{H}_\text{tot} = 0.
\end{equation}
Now consider the random fluctuations of both gravitational fields and scalar fields. We assume that such field fluctuations are completely local and independent from each other. The variance of the fluctuations of gravitational fields is still given by \eqref{varianceW}. Denote the fluctuation of the scalar fields as $\xi=\Delta\phi$, the variance of the fluctuations of the scalar fields is \cite{Yang2025}
\begin{equation}
\label{varianceXi}
    \langle\xi(x)\xi(y)\rangle=\frac{\hbar\Delta t}{2}\delta(x-y).
\end{equation}
For the ensemble of field configurations on a hypersurface $\Sigma_t$ with field variables $h$ and $\phi$, we introduce the probability density $\rho[h,\phi, t]$. Define the information metrics for the total relative entropy accumulated over a period of time label from $t_0$ to $t_{N-1}$ as
\begin{equation}
    I_f = \sum_{i=0}^{N-1}\langle D_\text{KL}(\rho[h,\phi, t_i]||\rho[h+w,\phi+\xi, t_i])\rangle_{w,\xi}.
\end{equation}
Following similar derivation steps in Appendix \ref{appendix:SE}, and using \eqref{varianceW} and \eqref{varianceXi}, we find that
\begin{equation}
\begin{split}    
    I_f =& \frac{\hbar}{2}\int \mathcal{D}h\mathcal{D}\phi d^3xdt\{\frac{16\pi GN}{\rho}\frac{\delta\rho}{\delta h_{ij}}\frac{G_{ijkl}}{\sqrt{h}}\frac{\delta\rho}{\delta h_{kl}}\\ &+\frac{N}{2\rho\sqrt{h}}(\frac{\delta\rho}{\delta\phi})^2\}.
\end{split}
\end{equation}
With this, the Lagrangian density of the quantum ensemble of the combined gravitational fields and scalar fields is
\begin{equation}
\begin{split}
    &\mathcal{L}_\text{q}[h,\phi]=\rho\dot{S}+\rho\int d^3x[N(\mathcal{H}_g+\mathcal{H_\phi})+N_i\mathcal{H}^i]\\
    &+\frac{\hbar^2}{4}\int d^3x\frac{N}{\sqrt{h}}\{\frac{16\pi G}{\rho}\frac{\delta\rho}{\delta h_{ij}}G_{ijkl}\frac{\delta\rho}{\delta h_{kl}}+\frac{1}{2\rho}(\frac{\delta\rho}{\delta\phi})^2\}.
\end{split}
\end{equation}
We impose similar constraints to the functional $S$ and $\rho$ such that they must be diffeomorphism invariant. Follow the framework of quantization with constraints as described in Section \ref{sec:WDE}, and choose the shift gauge of $N_i=0$, we obtain two functional derivative equations.
\begin{align}
\label{GFCont}
    32\pi G\frac{\delta }{\delta h_{ij}}(\rho \frac{G_{ijkl}}{\sqrt{h}}\frac{\delta S}{\delta h_{kl}})+\frac{\delta}{\delta\phi}(\frac{\rho}{\sqrt{h}}\frac{\delta S}{\delta\phi})&=0,\\
    \label{GFHJE}
    \mathcal{H}_g+\mathcal{H}_\phi + Q_g+Q_\phi &= 0,
\end{align}
where
\begin{align}
    Q_g[\rho]&=-\frac{16\pi G\hbar^2}{\sqrt{\rho}} \frac{\delta }{\delta h_{ij}}( \frac{G_{ijkl}}{\sqrt{h}}\frac{\delta \sqrt{\rho}}{\delta h_{kl}}), \\
    Q_\phi[\rho]&= -\frac{\hbar^2}{2\sqrt{h}}\frac{1}{\sqrt{\rho}} \frac{\delta^2 \sqrt{\rho}}{\delta \phi^2}.
\end{align}
$Q_g$ and $Q_\phi$ are the Bohm quantum potentials due to gravitational field fluctuations and scalar field fluctuations, respectively. 

We follow the strategy proposed in \cite{HallBook} to rewrite the functional $S$ and $\rho$ in the following forms,
\begin{equation}
\label{separatePsi}
\begin{split}
    S[h,\phi]&=S_0[h]+S_1[h,\phi],\\
    \rho [h,\phi]&=\rho_0[h]\rho_1[h,\phi],
\end{split}
\end{equation}
and assume that the pair $(\rho_0, S_0)$, when written in wave functional form, satisfies the Wheeler-DeWitt equation of pure gravitational fields\footnote{Effectively, in terms of wave functional $\Psi=\sqrt{\rho}e^{iS/\hbar}$, \eqref{separatePsi} is equivalent to rewrite $\Psi$ as $\Psi_0[h]\Psi_1[h,\phi]$.}. The idea here is that by substituting them into \eqref{GFCont} and \eqref{GFHJE}, we wish to separate the dynamics between the gravitational fields and the scalar fields. 

First, we can verify that $Q_\phi[\rho]=Q_\phi[\rho_1]$, and 
\begin{equation}
\label{Qg}
    Q_g[\rho]=Q_g[\rho_0]+Q_g[\rho_1]-8\pi G\hbar^2\frac{\delta\ln\rho_0}{\delta h_{ij}}\frac{G_{ijkl}}{\sqrt{h}}\frac{\delta\ln\rho_1}{\delta h_{kl}}.
\end{equation}
Applying the identity in \eqref{BohmP}, we expand $Q_g[\rho_0]$ as
\begin{equation}
\label{Qg0}
    Q_g[\rho_0] =4\pi G\hbar^2(\frac{\delta \ln\rho_0}{\delta h_{ij}}\frac{G_{ijkl}}{\sqrt{h}}\frac{\delta \ln\rho_0}{\delta h_{kl}}-\frac{2}{\rho_0}\frac{\delta }{\delta h_{ij}}\frac{G_{ijkl}}{\sqrt{h}}\frac{\delta \rho_0}{\delta h_{kl}})
\end{equation}
If we introduce the assumption
\begin{equation}
\label{approx1}
    |\frac{\delta\ln\rho_0}{\delta h_{ij}}| \gg |\frac{\delta\ln\rho_1}{\delta h_{ij}}|,
\end{equation}
the third term in \eqref{Qg} can be ignored compared to the first term in \eqref{Qg0}. This leaves
\begin{equation}
    Q_g[\rho]\approx Q_g[\rho_0]+Q_g[\rho_1].
\end{equation}
Second, the rate equation for the gravitational fields can be obtained from \eqref{extCurv}
\begin{equation}
    \dot{h}_{ij}=2NK_{ij}+D_iN_j+D_jN_i.
\end{equation}
In the shift gauge, $N_i=0$. Combined with the definition of $\pi^{ij}$ in \eqref{momentumConj} and \eqref{momentumConj2}, we find that
\begin{equation}
    \dot{h}_{ij} = \frac{32\pi GN}{\sqrt{h}}G_{ijkl}\frac{\delta S}{\delta h_{kl}}.
\end{equation}
With these definitions, \eqref{GFHJE} is broken down to
\begin{equation}
\label{GFHJE2}
\begin{split}
    &\{\mathcal{H}_g[S_0]+Q_g[\rho_0]\}+\{\mathcal{H}_\phi[S_1]+Q_\phi[\rho_1] \\&+\frac{1}{N}\frac{\delta S_1}{\delta h_{ij}}\dot{h}_{ij}-\frac{16\pi G}{\sqrt{h}}\frac{\delta S_1}{\delta h_{ij}}G_{ijkl}\frac{\delta S_1}{\delta h_{kl}}+Q_g[\rho_1]\}=0.
\end{split}
\end{equation}
Next, we break down \eqref{GFCont} to
\begin{equation}
\label{GFCont2}
\begin{split}
      &32\pi G\rho_1\frac{\delta }{\delta h_{ij}}\{\rho_0 \frac{G_{ijkl}}{\sqrt{h}}(\frac{\delta S_0}{\delta h_{kl}}+\frac{\delta S_1}{\delta h_{kl}})\}\\
      &+\rho_0\{\frac{1}{N}\frac{\delta \rho_1}{\delta h_{ij}}\dot{h}_{ij}+\frac{\delta}{\delta\phi}(\frac{\rho_1}{\sqrt{h}}\frac{\delta S_1}{\delta\phi})\}=0.
\end{split}
\end{equation}
Similarly to \eqref{approx1}, we assume
\begin{equation}
    \label{approx2}
        |\frac{\delta S_0}{\delta h_{ij}}| \gg |\frac{\delta S_1}{\delta h_{ij}}|.
\end{equation}
Both assumptions in \eqref{approx1} and \eqref{approx2} imply that the dependence on the gravitational variables is much stronger for $(\rho_0, S_0)$ than for $(\rho_1, S_1)$. In other words, $(\rho_0, S_0)$ account for most of the physical effects of gravitational degrees of freedom, while $(\rho_1, S_1)$ are considered as correction terms required by the presence of quantized scalar fields. With \eqref{approx2}, \eqref{GFCont2} becomes
\begin{equation}
\label{GFCont3}
\begin{split}
      &32\pi G\rho_1\frac{\delta }{\delta h_{ij}}\{\rho_0 \frac{G_{ijkl}}{\sqrt{h}}\frac{\delta S_0}{\delta h_{kl}}\}\\&+\rho_0\{\frac{1}{N}\frac{\delta \rho_1}{\delta h_{ij}}\dot{h}_{ij}+\frac{\delta}{\delta\phi}(\frac{\rho_1}{\sqrt{h}}\frac{\delta S_1}{\delta\phi})\}=0.
\end{split}
\end{equation}
Since the pair $(\rho_0, S_0)$ satisfies the equations in \eqref{var11} and \eqref{var12}, we further simplify \eqref{GFHJE2} and \eqref{GFCont3} as
\begin{align}
\label{GFHJE4}
    &\frac{\delta S_1}{\delta h_{ij}}\dot{h}_{ij}+N\{\mathcal{H}_\phi[S_1]+Q_\phi[\rho_1]+\Gamma [\rho_1,S_1]\}=0\\
    \label{GFCorr}
    &\Gamma [\rho_1,S_1]=-\frac{16\pi G}{\sqrt{h}}\frac{\delta S_1}{\delta h_{ij}}G_{ijkl}\frac{\delta S_1}{\delta h_{kl}}+Q_g[\rho_1]\\
    \label{GFCont4}
    &\frac{\delta \rho_1}{\delta h_{ij}}\dot{h}_{ij}+N\frac{\delta}{\delta\phi}(\frac{\rho_1}{\sqrt{h}}\frac{\delta S_1}{\delta\phi})=0.
\end{align}
The term $\Gamma [\rho_1,S_1]$ is considered a correction term.

The time parameter $t$ in the rate equation of the gravitational fields is a label to track how the gravitational fields change along the slicing of the hypersurfaces. We read it off as a physical clock to track the changes of the functional $S_1$ and $\rho_1$. Thus, by the chain rule of functional derivative\footnote{Mathematically, since $S_1$ is also a functional of $\phi$, there must be another term $\int d^3x (\delta S_1/\delta \phi)\dot{\phi}$. However, we choose to read off time from the rate equation of gravitational fields only, and this term is discarded.},
\begin{equation}
\label{ETime}
\begin{split}
    \frac{\partial S_1}{\partial t}&=\int d^3x \frac{\delta S_1}{\delta h_{ij}}\dot{h}_{ij},\\
    \frac{\partial \rho_1}{\partial t}&=\int d^3x \frac{\delta \rho_1}{\delta h_{ij}}\dot{h}_{ij}.
\end{split}
\end{equation}
Taking the integration of \eqref{GFHJE4} and \eqref{GFCont4} with respect to the spatial variable $x$ and applying \eqref{ETime}, we get
\begin{align}
    &\frac{\partial S_1}{\partial t}+\int d^3xN\{\mathcal{H}_\phi[S_1]+Q_\phi[\rho_1]+\Gamma [\rho_1,S_1]\}=0,\\
    &\frac{\partial \rho_1}{\partial t}+\int d^3x \frac{N}{\sqrt{h}}\frac{\delta}{\delta\phi}(\rho_1\frac{\delta S_1}{\delta\phi}) =0.
\end{align}
Define a wave functional $\Psi_1[h,\phi, t]=\sqrt{\rho_1}e^{iS_1/\hbar}$, the above two equations are combined into a non-linear Schr\"{o}dinger equation for $\Psi_1$,
\begin{equation}
    \label{SEPhi}
    \begin{split}
        i\hbar\frac{\partial\Psi_1}{\partial t}=\int d^3x N[-\frac{\hbar^2}{2\sqrt{h}}\frac{\delta^2}{\delta\phi^2}+\frac{\sqrt{h}}{2}h^{ij}\partial_i\phi\partial_j\phi+\Gamma]\Psi_1.
    \end{split}
\end{equation}
The non-linear correction term $\Gamma$ in \eqref{GFCorr} contains two sub-terms. The first correction sub-term arises from the presence of classical gravitational fields and can be explicitly expressed as
\begin{equation}
    \label{GFCorr1}
    \begin{split}
    \Gamma_\text{cl}&=-\frac{16\pi G}{\sqrt{h}}\frac{\delta S_1}{\delta h_{ij}}G_{ijkl}\frac{\delta S_1}{\delta h_{kl}}\\
    &=-4\pi G\hbar^2\frac{\delta \ln(\Psi_1/\Psi^*_1)}{\delta h_{ij}}\frac{G_{ijkl}}{\sqrt{h}}\frac{\delta \ln(\Psi_1/\Psi^*_1)}{\delta h_{kl}}.
    \end{split}
\end{equation}
The second correction sub-term is due to the quantum effect of the gravitational fields since it originates from the information metrics of field fluctuations, 
\begin{equation}
    \label{GFCorr2}
    \begin{split}
    \Gamma_\text{q}&=Q_g[\rho_1]\\
    &=-\frac{16\pi G\hbar^2}{(\Psi_1\Psi^*_1)^{1/4}}\frac{\delta }{\delta h_{ij}}\{\frac{G_{ijkl}}{\sqrt{h}}\frac{\delta (\Psi_1\Psi^*_1)^{1/4}}{\delta h_{kl}}\}.
    \end{split}
\end{equation}
Both correction terms are of the order $\mathcal{O}(G\hbar^2)$, much smaller than the potential term $h^{ij}\partial_i\phi\partial_j\phi$ in \eqref{SEPhi}. If $\Gamma$ is ignored, \eqref{SEPhi} is reduced to the regular Schr\"{o}dinger equation of the wave functional for the scalar fields, as expected. The quantum correction term $\Gamma_\text{q}$ is not presented in the results of \cite{HallBook}. In the energy eigenstate of scalar fields coupled with gravitational fields, e.g., the ground state of equation \eqref{SEPhi}, we have $S_1=Et$ where $E$ is the energy level of the eigenstate. Then $\Gamma_\text{cl}=0$ and $\Gamma_\text{q}\ne 0$. In such conditions, $\Gamma_\text{q}$ is the correction term to the regular Schr\"{o}dinger equation that is purely due to the quantum effect of gravitational fields. This may provide a mechanism for detecting the quantum effect of gravitational fields, which has been a highly interested subject of recent research on quantum gravity~\cite{Vedral,Vedral2}.

\section{Discussion and conclusions} 
\label{sec:discussion}

\subsection{Comparisons with Hall and Reginatto's Framework}
In Step III of the quantization framework described in Section~\ref{sec:LIP}, we adopt the ensemble description of field configurations and treat the pair of functionals $(\rho, S)$ as generalized canonical variables, where $S$ plays the role of a conjugate momentum functional. This ensemble-based formulation was introduced by Hall and Reginatto~\cite{HallBook} and provides a consistent mathematical structure to describe classical ensembles in the configuration space. Their framework establishes a useful foundation for extending classical statistical descriptions toward quantum theory. 

It is important to emphasize that in the present work we adopt this ensemble formulation only at the level of classical field representations in Step III. The quantization procedure introduced in Steps II and IV is conceptually distinct from that in~\cite{HallBook}. In their approach, quantization is based on an “exact uncertainty” principle, in which momentum fluctuations are directly constrained by position uncertainty. By contrast, the present framework introduces quantum effects through an extended stationary action principle, in which an additional contribution to the action arises from information-theoretic metrics characterizing field fluctuations. These metrics are constructed from relative entropy measures that quantify the information distance between probability distributions with and without fluctuations.

From this perspective, the exact uncertainty principle encodes quantum fluctuations through a direct relationship between conjugate variables, whereas the present approach encodes fluctuations through a variational information functional added to the classical action. This reformulation provides a different conceptual route to quantization, in which the information-theoretic structure plays a central role.

In the treatment of emergent time~\cite{HallBook}, Hall and Reginatto consider a hybrid framework in which gravitational fields remain classical while matter fields are quantized, leading to a semiclassical gravity–quantum matter system. In contrast, the present work does not impose such a restriction. Both gravitational and matter fields are treated within the same quantization framework. As shown in Section~\ref{sec:etime}, this allows the derivation of a Schrödinger equation for the scalar field using an emergent time parameter defined from the dynamical evolution of the gravitational field ensemble. In this fully quantized setting, additional correction terms appear in Eq.~\eqref{SEPhi}, which are absent in the semiclassical treatment of~\cite{HallBook}.

\subsection{Comparisons with Standard Second Quantization Frameworks}
The standard formulations of second quantization in quantum field theory—namely canonical quantization and the path integral approach—together with the framework developed in this work, all originate from the underlying Lagrangian formalism. Among these, the path integral formulation is often regarded as the most direct construction. However, it implicitly relies on the existence of a linear Schrödinger equation for the wave functional, with the path integral and Schrödinger formulations being formally equivalent under appropriate assumptions~\cite{FeynmanBook}. This raises a natural question: whether the path integral approach can be consistently extended to systems whose effective dynamics lead to a non-linear Schrödinger equation, such as Eq.~\eqref{SEPhi}. 

Both canonical quantization and the present framework begin by identifying conjugate momenta from the classical Lagrangian. However, a key distinction arises in the subsequent step. In canonical quantization, the classical field variables and their conjugate momenta are elevated to operators as a fundamental postulate, which introduces ambiguities such as operator ordering, as exemplified in the Wheeler–DeWitt equation derived within the canonical approach. In contrast, the framework developed in this work does not require a postulated operator promotion. Instead, operator structures emerge only after the variational quantization procedure is complete, where they serve as mathematical representations of the resulting dynamics. This feature avoids operator-ordering ambiguity at the fundamental formulation level. Furthermore, in standard canonical quantization, the quantized Hamiltonian is assumed to generate a linear Schrödinger evolution in the wave functional representation. However, this assumption is not generally guaranteed once additional structural contributions are present, as illustrated by Eq.~\eqref{SEPhi}, where the effective evolution equation becomes non-linear.

Alternative approaches to quantum gravity, including string theory and loop quantum gravity, provide different strategies for addressing some of these foundational issues. Nevertheless, many of these approaches still rely, either explicitly or implicitly, on the promotion of classical variables and their conjugate momenta to operators satisfying commutation relations or on path integral formalism.

\subsection{Implications of Relative Entropy}
The advantages of using the relative entropy as fundamental information metrics compared to Fisher information have been summarized in Section \ref{sec:IF}. It allows us to potentially expand the definition of $I_f$ using another generalized form of relative entropy such as R\'{e}nyi divergence or Tsallis divergence, as shown in Section \ref{sec:Tsallis}. To derive \eqref{FisherInfo2}, one needs the identity \eqref{varianceW}, which can be derived through
the same extended principle of least action, using the relative entropy again, in a short time step, as shown in Section \ref{sec:shortTimeStep}. This makes the framework mathematically quite elegant.

There is another possible implication, which is more subtle. In modern quantum gravity theory, the holographic correspondence is an interesting clue to understanding quantum gravity. The holographic correspondence states that the relative entropy of a region on the boundary surface equals the relative entropy of the bulk volume correspondence to the boundary surface~\cite{FLM,JLMS}. Schematically, this statement can be mathematically expressed as 
\begin{equation}
    S_\text{rel}^\text{CFT}(A) = S_\text{rel}^\text{Bulk}(W_A).
\end{equation}
Here $S_\text{rel}^\text{CFT}(A)$ is the relative entropy of a surface region $A$ described by a conformal field theory without gravity, while $S_\text{rel}^\text{bulk}(W_A)$ is the relative entropy of the bulk volume $W_A$ (with the boundary surface $A$) in certain types of spacetime such as an Anti-de Sitter (AdS) spacetime and described by a quantum theory with gravity. We may wonder if we can interpret the relative entropy defined in \eqref{FisherInfo2} as $S_\text{rel}^\text{bulk}(W_A)$. The relative entropy in \eqref{FisherInfo} is defined in the bulk of spacetime with gravity as the information distance between probability density with and without field fluctuations. One can also define another relative entropy in the boundary surface without gravity and again as the information distance between probability density with and without field fluctuations. Then, it is interesting to investigate whether there is a connection between the two kinds of relative entropy. 


\subsection{Limitations}
The assumption of field fluctuations serves as the foundation for defining the information metric $I_f$, which ultimately gives rise to the quantum behavior of gravitational fields. However, we do not provide a concrete physical model for these fluctuations. The underlying physics governing field fluctuations is expected to be complex and may hold the key to a deeper understanding of quantum field theory. Exploring this in detail is beyond the scope of this paper. Our goal here is to minimize the number of assumptions required to derive the Wheeler-DeWitt equation for the wave functional of the gravitational fields, allowing future research to focus on justifying and refining these assumptions. 

The emergent time described in Section \ref{sec:etime} attempts to solve the problem of time using a classical clock from the rate equation of the gravitational fields. It is a non-perturbative approach, as compared to the WKB approximation, which is a perturbative solution. A notable full quantum treatment to solve the problem of time is the Page-Wootters (PW) formalism~\cite{PW}, where time arises from quantum correlations inside a globally timeless state. In the PW framework, a quantum clock state is defined first, then the physical state of a matter field is a projection of the timeless state to the clock state. In a sense, the physical state of matter fields is a quantum state conditioned on the clock state. Time is relational in this formalism, there is no absolute time. The problem of time is still an active research topic in quantum gravity. A comprehensive review can be found in \cite{Anderson}.

\subsection{Conclusions}
In this paper, we have developed a variational framework for the quantization of gravitational fields based on an extended stationary action principle. Originally introduced in the context of non-relativistic quantum theory~\cite{Yang2023} and subsequently applied to scalar and fermionic fields, this principle provides a unified perspective on the transition from classical to quantum field theory. Within the present work, and in alignment with the ADM formulation of general relativity, the framework is extended to constrained systems, leading to a derivation of the Wheeler–DeWitt equation for the gravitational wave functional within this formalism. Furthermore, for gravitational fields coupled to a scalar field, we show that an emergent time parameter can be defined through the rate equation of the gravitational fields. This construction allows one to recover a Schrödinger equation for the scalar-field wave functional, supplemented by nonlinear correction terms arising from the coupling to the quantum gravitational degrees of freedom.

The extended stationary action principle introduces an information-theoretic structure into the Lagrangian formulation of field theory. As described in Section~\ref{sec:LIP}, the framework is based on two central assumptions. First, a relative entropy functional is introduced to quantify additional information associated with field fluctuations, thereby defining an information-theoretic metric on the space of configurations. Second, the Planck constant sets the minimal scale at which such fluctuations contribute to observable dynamics, allowing the entropy-based contribution to enter as a correction to the classical action. In the limit where this contribution becomes negligible, the formalism reduces to classical field theory. In this way, quantum behavior emerges from the incorporation of information-theoretic corrections into the variational principle.

From a structural standpoint, the framework provides an alternative to both canonical quantization and path integral formulations. A key feature is that the constraint structure of general relativity is incorporated directly into the variational principle, so that quantization and constraint enforcement are implemented simultaneously. This avoids the ordering ambiguity inherent in the Dirac and reduced approaches and leads to a formulation in which the quantum constraints are satisfied by construction.

The results obtained here, together with those in Refs.~\cite{Yang2023, Yang2024, Yang2025, Yang2026}, illustrate the flexibility and broad applicability of the extended stationary action principle across non-relativistic quantum mechanics and relativistic quantum field theories. While the present framework provides a coherent treatment of operator ordering, constrained quantization, and aspects of the problem of time, further work is required to clarify its implications for renormalization and ultraviolet behavior. 

Finally, the introduction of relative entropy as a foundational element suggests potential connections to modern developments in quantum gravity. In particular, it is of interest to investigate whether the relative entropy functional employed here is related to the notion of relative entropy appearing in holographic dualities between conformal field theories and bulk gravitational systems~\cite{Takayanagi, FLM, JLMS}. Exploring this connection may provide further insight into the role of information-theoretic structures in quantum gravity.



\section*{Data Availability and Funding Declaration}
Data supporting the findings of this study are available in the article. Funding information - not applicable.





\onecolumngrid

\pagebreak

\appendix

\section{Quantization of Scalar Fields}
\label{appdx:qmscalar}
In this Appendix, we review the quantization of massive scalar fields. Although the theory has been developed previously~\cite{Yang2024}, our purpose here is to explicitly illustrate the five-step procedure outlined in Section II.

First, the Lagrangian of massive scalar fields is
\begin{equation}
    \label{LD}
        \mathcal{L} = \frac{1}{2}[\partial_{\mu}\phi(x)]^2 - \frac{1}{2}m^2[\phi(x)]^2 =\frac{1}{2}[\dot{\phi}(x)]^2 - \frac{1}{2}([\nabla\phi(x)]^2+m^2[\phi(x)]^2).
\end{equation}
The momentum conjugate is $\pi(x) = \frac{\partial\mathcal{L}}{\partial(\partial_0\phi)}= \dot{\phi}(x)$. The Hamiltonian is constructed by a Legendre transform of the Lagrangian~\cite{Long}
\begin{equation}
    \label{Hamiltonian}
    H[\phi,\pi] = \int d^3{x}\{\pi(x)\dot{\phi}(x) - \mathcal{L}\} = \int d^3 {x} \{\frac{1}{2}[\dot{\phi}(x)]^2 + V\}.
\end{equation}
In Step II, we consider the field fluctuations in equal times hyper-surfaces for an infinitesimal time internal $\Delta t$. At a given time $t\to t+\Delta t$ in the hyper-surface $\Sigma_t$, the field configuration fluctuates randomly, $\phi\to \phi + \omega$, where $\omega=\Delta\phi$ is the change of field configuration due to random fluctuations. Define the probability that the field configuration will transition from $\phi$ to $\phi + \omega$ as $p[\phi + \omega|\phi]\mathcal{D}\omega$. The expectation value of the classical action over all possible field fluctuations is $S_c=\int p[\phi + \omega|\phi]\mathcal{L}d^3 {x}\mathcal{D}\omega dt$ where $\mathcal{L}$ is given by (\ref{LD}) for a scalar field. For an infinitesimal time internal $\Delta t$, one can approximate $\dot\phi=\Delta\phi/\Delta t=\omega/\Delta t$. The classical action for the infinitesimal time internal $\Delta t$ is approximately given by 
\begin{equation}
\label{action1}
    S_c=\int p[\phi + \omega|\phi]\mathcal{D}\omega\int_{\Sigma_t}\{\frac{[\omega(x)]^2}{2\Delta t}+V(\phi(x))\Delta t\}d^3 {x}.
\end{equation}
The information metrics $I_f$ are supposed to capture the additional revelation of information due to field fluctuations in $\Sigma_t$. Thus, it is naturally defined as a relative entropy, or more specifically, the Kullback–Leibler divergence, to measure the information distance between $p[\phi + \omega|\phi]$ and some prior probability distribution. Since field fluctuations are completely random, it is intuitive to assume the prior distribution with maximal ignorance~\cite{Caticha2019, Jaynes}. That is, the prior probability distribution is a uniform distribution $\sigma$. Therefore,
\begin{align*}
    I_f  &:= D_{KL}(p[\phi + \omega|\phi]|| \sigma) = \int p[\phi + \omega|\phi]ln[p[\phi + \omega|\phi]/\sigma]\mathcal{D}\omega.
\end{align*}
Combined with (\ref{action1}), the total action functional defined in (\ref{totalAction}) is
\begin{align*}
    S_t = &\int p[\phi + \omega|\phi]\mathcal{D}\omega\int (\frac{[\omega(x)]^2}{2\Delta t} + V(\phi(x)) \Delta t)d^3 {x} \\
        &+ \frac{\hbar}{2}\int p[\phi + \omega|\phi]ln[p[\phi + \omega|\phi]/\sigma]\mathcal{D}\omega.
\end{align*}
Taking the variation $\delta S_t = 0$ with respect to $p$, one obtains
\begin{equation}
\label{transP}
    p[\phi + \omega|\phi] = \frac{1}{Z}e^{-\frac{1}{\hbar\Delta t}\int[\omega(x)]^2 d^3{x}}
\end{equation}
where $Z$ is a normalization factor. Equation (\ref{transP}) shows that the transition probability density is a Gaussian distribution. The variance is
\begin{equation}
\label{variance}
    \langle \omega(x)\omega(x')\rangle = \frac{\hbar\Delta t}{2}\delta ({x}-{x'}).
\end{equation}

In the third step, we apply the canonical transformation in classical field theory by introducing a pair of generalized canonical coordinates $(\Phi, \Pi)$. Denote $K(\Phi, \Pi)$ and $L'(\Phi, \Pi)$ the Hamiltonian and the Lagrangian in the generalized coordinate system. They are related via $L'= \Pi\dot{\Phi}-K$. The extended canonical transformation gives the following results,
\begin{align}
    \label{type12}
    L' &=\frac{\partial S}{\partial t} +H, \\
    \pi(x) &= \frac{\delta S}{\delta\phi(x)},
\end{align}
where $S[\phi,\pi]$ is a generating functional. The action integral in the generalized canonical coordinates becomes
\begin{equation}
    \label{extActionA}
    A_c = \int^{t_B}_{t_A}dt L' = \int^{t_B}_{t_A}dt \{\frac{\partial S}{\partial t} + H(\phi, \frac{\delta S}{\delta \phi})\}.
\end{equation}

In order to describe the time evolution of the systems that are subject to uncertainty, we adopt the theoretical framework of ensemble in configuration space, developed by Hall and Reginatto~\cite{HallBook}. In this framework, the uncertainty of field configurations is described by an esemble of field configurations with a probability density over the configuration space, $\rho(\phi, t)$.  The Lagrangian density for the ensemble is naturally defined as $\mathcal{L}=\rho L'$. Thus, the expectation value of the classical action of the ensemble is
\begin{equation}
    \label{extActionE}
    A_e = \int \mathcal{D}\phi dt \mathcal{L} = \int \mathcal{D}\phi dt \rho \{\frac{\partial S}{\partial t} + H(\phi, \frac{\delta S}{\delta \phi})\}.
\end{equation}
As shown by Hall and Reginatto~\cite{HallBook,Hall:2001}, the Hamilton–Jacobi and continuity equations can be obtained from the classical action $A_e$ by fixed-point variation with respect to $\rho$ and $S$, respectively. In particular, taking variation of $A_e$ with respect to $\rho$ yields the Hamilton–Jacobi equation,
\begin{equation}
    \label{HJE}
    \frac{\partial S}{\partial t }+ \frac{1}{2}\int d^3\{(\frac{\delta S}{\delta\phi( x)})^2 + V[\phi]\} = 0.
\end{equation}
Taking variation of $A_e$ with respect to $S$ gives the continuity equation 
\begin{equation}
\label{contEqq}
   \frac{\partial \rho}{\partial t} + \int \frac{\delta}{\delta\phi(x)}(\rho\frac{\delta S}{\delta\phi(x)}) d^3\mathbf{x} = 0.
\end{equation}
These results suggest that $(\rho, S)$ can also be considered as a pair of generalized canonical coordinates for the ensemble. To confirm this observation, first, we notice that the Lagrangian density of the ensemble is
\begin{equation}
    \mathcal{L}_e[\rho, S] = \rho \{\frac{\partial S}{\partial t} + H(\phi, \frac{\delta S}{\delta \phi})\}.
\end{equation}
It follows that
\begin{equation}
    \rho = \frac{\delta \mathcal{L}_e}{\delta(\partial_t S)}
\end{equation}
Thus, $\rho$ can be considered as the generalized conjugate momentum for $S$. Second, the classical Hamiltonian density and the total Hamiltonian of the ensemble are
\begin{equation}
    \label{eHamiltonian}
    \begin{split}
    \mathcal{H}_e[\rho, S] &= \rho\dot{S}-\mathcal{L} = -\rho H(\phi, \frac{\delta S}{\delta \phi}),\\
    H_e[\rho, S] &= \int dx \mathcal{H}_e= -\int dx \rho H(\phi, \frac{\delta S}{\delta \phi}).
    \end{split}
\end{equation}
We can verify that equations \eqref{HJE} and \eqref{contEqq} are equivalent to the canonical equations based on $H_e$,
\begin{equation}
   \frac{\partial S}{\partial t}=-\frac{\delta H_e}{\delta \rho} , \textbf{ }\frac{\partial\rho}{\partial t}=\frac{\delta H_e}{\delta S}.
\end{equation}
These results justify the choice of $(\rho, S)$ as a pair of generalized canonical coordinates that, together with the Hamiltonian ensemble $H_e$, generates the dynamic equations of the classical ensemble. Eq. \eqref{extActionE} is the starting point for the quantization process in the next step.

In Step IV, to define the information metrics for the field fluctuations, $I_f$, we slice the time duration $t_A\to t_B$ into $N$ short time steps $t_0=t_A, \ldots, t_j, \ldots, t_{N-1}=t_B$, and each step is an infinitesimal period $\Delta t$. In an infinitesimal time period at time $t_j$, the scalar fields not only evolve according to the Hamilton-Jacobi equation, but also experience random fluctuations. Such an additional revelation of distinguishability due to the vacuum fluctuations on top of the classical trajectory is measured by the following definition,
\begin{align}
\label{DLDivergence}
    I_f := \sum_{i=0}^{N-1}\langle D_{KL}(\rho[\phi, t_i] || \rho[\phi+\omega, t_i])\rangle_{\omega} 
    =\sum_{i=0}^{N-1}\int \mathcal{D}\omega p[\omega] \int \mathcal{D}\phi \rho [\phi, t_i]ln \frac{\rho[\phi, t_i]}{\rho [\phi+\omega, t_i]}.
\end{align}
When $\Delta t\to 0$, and with the identity \eqref{variance}, $I_f$ turns out to be~\cite{Yang2023}
\begin{equation}
\label{FisherInfo}
    I_f = \frac{\hbar}{4}\int d^3{x} \mathcal{D}\phi dt \frac{1}{\rho[\phi, t]}(\frac{\delta\rho[\phi, t]}{\delta\phi(x)})^2 .
\end{equation}
Eq. (\ref{FisherInfo}) contains the term related to Fisher information~\cite{Frieden, Reginatto} for the probability density $\rho$. 
Inserting (\ref{extActionE}) and (\ref{FisherInfo}) into (\ref{totalAction}), we obtain the total action
\begin{equation}
    \label{totalAction2}
    A_t =  \int d^3{x}\mathcal{D}\phi dt\rho\{\frac{\partial S}{\partial t} + \int [\frac{1}{2}(\frac{\delta S}{\delta\phi(x)})^2 + V(\phi(x))  + \frac{\hbar^2}{8}(\frac{1}{\rho}\frac{\delta\rho}{\delta\phi(x)})^2 ]\}.
\end{equation}
Effectively, including $I_f$ amounts to adding the Fisher information term $F[\rho]$ to the Lagrangian density
\begin{equation}
    \mathcal{L}_q[\rho, S] = \rho(\dot{S}+H[x, \frac{\partial S}{\partial x}] + F[\rho]), \text{where } F[\rho]=\frac{\hbar^2}{8}(\frac{1}{\rho}\frac{\delta\rho}{\delta\phi(x)})^2.
\end{equation}

In step V, we perform the variation procedure on $A_t$ with respect to $S$, which gives the continuity equation. On the other hand, performing the variation with respect to $\rho$ leads to the quantum Hamilton-Jacobi equation,
\begin{align}
\label{QHJ}
    \frac{\partial S}{\partial t} =- \int\{\frac{1}{2}(\frac{\delta S}{\delta\phi(x)})^2 + V(\phi(x)) - \frac{\hbar^2}{2R}\frac{\delta^2 R}{\delta\phi^2(x)} \}d^3{x},
\end{align}
where $R[\phi, t]=\sqrt{\rho[\phi, t]}$. The last term in the R.H.S. of (\ref{QHJ}) is the scalar field equivalence of the Bohm quantum potential~\cite{Bohm1952}. In non-relativistic quantum mechanics, the Bohm potential is considered responsible for the non-locality phenomenon in quantum mechanics~\cite{Bohm2}. Its origin is mysterious. Here we show that it originates from information metrics related to relative entropy, $I_f$. 

Defining a complex functional $\Psi[\phi,t]=R[\phi, t]e^{iS[\phi, t]/\hbar}$, the continuity equation and the extended Hamilton-Jacobi equation (\ref{QHJ}) can be combined into a single functional derivative equation.
\begin{equation}
    \label{SE}
    i\hbar\frac{\partial\Psi[\phi, t]}{\partial t} = \{\int[-\frac{\hbar^2}{2}\frac{\delta^2}{\delta\phi^2(x)} + V(\phi(x))]d^3{x}\}\Psi[\phi, t].
\end{equation}
This is the Schr\"{o}dinger equation for the wave functional $\Psi[\phi,t]$

Once the Schr\"{o}dinger equation for the wave functional is derived and the correct Hamiltonian operator is identified, one can return to the standard operator-based approach. For instance, operators for particle creation and annihilation can be defined, and the energy of the ground state or excited states can be calculated. The important point here is that the Schr\"{o}dinger equation is derived from the first principle rather than through a postulate in standard canonical quantization.

\section{Constrained Ensemble Systems}
\label{appdx:qconst}
With the variational formulation presented in Appendix \ref{appdx:qmscalar}, the incorporation of constraints is straightforward: they can be incorporated into the Lagrangian through Lagrange multipliers in Step IV, and the variation procedure is carried out as usual. 

We can first examine a classical ensemble with constraints. Denote the primary constraints by $\{\Phi_\alpha\}$. First, we need to express these primary constraints in terms of $\rho$ and $S$ after the canonical transformation in Step III. Then, the constraints are added to the Lagrangian density for the constrained ensemble. 
\begin{equation}
\label{Lqc}
    \mathcal{L}_\text{en}[\rho, S,\lambda] = (\rho\dot{S}+\rho H[x, \frac{\partial S}{\partial x}])+\lambda_\alpha\Phi^\alpha[\rho, S],
\end{equation}
where $\lambda_\alpha$ are Lagrange multipliers and the convention on summation over $\alpha$ is assumed. The total action with constraints is
\begin{equation}
    A_t = \int dtdx \mathcal{L}_\text{en}[\rho, S, \lambda].
\end{equation}
The total Hamiltonian of the quantum ensemble, $H_T$, is defined by
\begin{equation}
    H_T \equiv \int dx (\rho\dot{S}-\mathcal{L}_\text{en})=-\int dx \{\rho H  + \lambda_\alpha\Phi^\alpha[\rho, S] \}
    =H_\text{en} - \int dx \lambda_\alpha\Phi^\alpha[\rho, S].
\end{equation}
To determine whether a constraint induces secondary constraints, we evaluate the Poisson bracket
\begin{equation}
    \dot{\Phi}_\alpha = \{\Phi_\alpha, H_T\}.
\end{equation}
The secondary constraint arises if the Poisson bracket does not vanish on shell. Note that there may be secondary constraints when the Hamiltonian is expressed in the $(x,p)$ phase space. But after the phase space is changed to $(\rho, S)$, we need to re-evaluate the secondary constraint. This reflects the fact that the total Hamiltonian of the ensemble is different after we introduce the probability density $\rho$ for the ensemble.

Next, we perform the variation procedure on $A_t$ with respect to $\rho, S$ and $\lambda_\alpha$, and solve the resulting equations to obtain the final equations of motion. Alternatively, we use the total Hamiltonian to derive the dynamic equation 
\begin{equation}
\label{canonicalEq}
    \dot{S}= \{S, H_T\}, \text{ }\dot{\rho}=\{\rho, H_T\}.
\end{equation}
If there are second class constraints, the Poisson bracket is replaced with the Dirac bracket \cite{Dirac},
\begin{equation}
\label{canonicalEq}
    \dot{S}= \{S, H_T\}_D, \text{ }\dot{\rho}=\{\rho, H_T\}_D.
\end{equation}

To extend the framework to a quantum ensemble, we just need to include the relative entropy term $I_f$ in the action functional, as shown in Section \ref{subsec:qconstraint}.


\section{Constraints for the Ensemble of Gravitational Fields}
\label{appdx:gravityconstraint}
First, the requirement that $S$ is invariant under time coordinate reparameterization, $\delta S/\delta t=0$, is equivalent to $H_{\text{ADM}}=0$, given the EHJ equation. This also leads to $N\mathcal{H}_\perp+N_i\mathcal{H}^i\approx0$.  For the ensemble of field configurations with probability density $\rho[h]$, we write this constraint as
\begin{equation}
    \label{C1}
    \mathcal{C}_1: \int \mathcal{D}h\text{ }\rho[h](\frac{\partial S}{\partial t})\approx0
\end{equation}
Next, we show the constraint induced by the requirement that $S$ is invariant under spatial transformation. Considering an infinitesimal diffeomorphism generated by $x^i\to x^i+\epsilon^i(x)$, we expect $\delta_\epsilon S =0$. The field metric is transformed as $h_{ij}\to h_{ij}-(D_i\epsilon_j + D_j\epsilon_i)$. Thus, $\delta h_{ij}=-(D_i\epsilon_j + D_j\epsilon_i)$, and
\begin{equation}
    \delta_\epsilon S = \int d^3x \frac{\delta S}{\delta h_{ij}}\delta h_{ij}=\int d^3x\{D_i(\frac{\delta S}{\delta h_{ij}})\epsilon_j+D_j(\frac{\delta S}{\delta h_{ij}})\epsilon_i\},
\end{equation}
where we applied integration by part in the second step. Since $\epsilon_i$ is arbitrary, we require $D_i(\frac{\delta S}{\delta h_{ij}})=0$. This is equivalent to the constraint $\mathcal{H}^i=0$. For the ensemble of field configurations with probability density $\rho[h]$, this form of constraint is written as the following form to be added to the Lagrangian
\begin{equation}
    \label{C2}
    \mathcal{C}^i_2: \int d^3x\mathcal{D}h\text{ }\rho[h]D_i(\frac{\delta S}{\delta h_{ij}})\approx0.
\end{equation}
Thus, $\mathcal{C}_1$ is equivalent to $N\mathcal{H}_\perp+N_i\mathcal{H}^i\approx0$, and $\mathcal{C}_2^i$ is equivalent to $\mathcal{H}^i\approx0$. According to Dirac's theory of constraints, $\mathcal{H}_\perp\approx0$ and $\mathcal{H}^i\approx0$ are first-class constraints. It follows that 
\begin{equation}
    \{\mathcal{C}_1, H_{\text{en}}\}\approx 0, \text{ }\{\mathcal{C}_2^i, H_{\text{en}}\}\approx 0, \text{ }\{\mathcal{C}_1, \mathcal{C}_2^i\}\approx 0,
\end{equation}
where $H_{\text{en}}$ is the Hamiltonian of the ensemble derived from \eqref{enH},
\begin{equation}
    H_{\text{en}} =\int \mathcal{D}h \text{ }\mathcal{H}_{\text{en}}= -\int \mathcal{D}h\text{ }\rho[h]\int d^3x(N\mathcal{H}_\perp[h_{ij},\frac{\delta S}{\delta h_{ij}}]+N_i\mathcal{H}^i[\frac{\delta S}{\delta h_{ij}}]).
\end{equation}
Therefore, no new constraint is induced from $\mathcal{C}_1$ and $\mathcal{C}_2^i$.

The constraint $\partial\rho/\partial t=0$ can be justified as follows. By the canonical equations,
\begin{equation}
    \dot{\rho}=\{\rho,H_{\text{en}}\} = \frac{\delta H_{\text{en}}}{\delta S}=-2\int d^3x \text{ }\frac{N}{\sqrt{h}} \frac{\delta }{\delta h_{ij}}\{\rho G_{ijkl}\frac{\delta S}{\delta h_{kl}}\},
\end{equation}
which is the continuity equation. On the other hand,
\begin{equation}
    \dot{\rho}=\{\rho,H_{\text{en}}\}=-\int \mathcal{D}h\text{ }\rho[h]\{\rho,\int d^3x(N\mathcal{H}_\perp[h_{ij},\frac{\delta S}{\delta h_{ij}}]+N_i\mathcal{H}^i[\frac{\delta S}{\delta h_{ij}}])\}.
\end{equation}
We expect $\rho[h]$ to be a physical observable. Thus, it must be invariant under the transformation generated by $\mathcal{H}_\perp$, that is, under the time reparameterization. It follows that the Poisson bracket $\{\rho, \mathcal{H}_\perp\}\approx 0$. Similarly, $\rho[h]$ must be invariant under the transformation generated by $\mathcal{H}^i$ and therefore $\{\rho, \mathcal{H}^i\}\approx 0$. Substituting them into the above equation leads to $\{\rho,H_{\text{en}}\} \approx 0$. Consequently, $\dot{\rho}\approx 0$. We rewrite this constraint in a form that can be added through the Lagrange multiplier:
\begin{equation}
    \label{C3}
    \mathcal{C}_3: \int \mathcal{D}h\text{ }\rho[h](\frac{\partial \rho}{\partial t})\approx0.
\end{equation}
The derivation of the constraint for $\rho[h]$ induced by invariance under spatial transformation is the same as that for $S$. The constraint is written as follows.
\begin{equation}
    \label{C4}
    \mathcal{C}^i_4: \int d^3x\mathcal{D}h\text{ }\rho[h]D_i(\frac{\delta \rho}{\delta h_{ij}})\approx0.
\end{equation}
Since $\mathcal{C}_3$ and $\mathcal{C}^i_4$ are induced by the gauge invariance of the spacetime differomorphism, they are also first-class constraints. They will not generate additional constraints. With $\mathcal{C}_1$ and $\mathcal{C}^i_2$, $\mathcal{C}_3$ and $\mathcal{C}^i_4$, all constraints are identified for the ensemble of gravitational fields.

\section{Information Metrics for Field Fluctuations}
\label{appendix:I_f}
Starting from the definition of the relative entropy in $I_f$, we expand the probability density $\rho[h+w]$ around $\rho[h]$.
\begin{align*}
    \ln\frac{\rho[h+w]}{\rho[h]} = & \ln \{1 + \frac{1}{\rho}\int d^3x\frac{\delta\rho}{\delta h_{ij}}w_{ij}\} 
     = \frac{1}{\rho}\int d^3x\frac{\delta\rho}{\delta h_{ij}}w_{ij} -\frac{1}{2\rho^2}[\int d^3x\frac{\delta\rho}{\delta h_{ij}}w_{ij}]^2
\end{align*}
Substitute the above expansion into (\ref{DLDivergence2}), and take the expectation values $\langle\cdot\rangle_{w}$. Using the fact that $\langle w_{ij}\rangle = 0$, and the identity in \eqref{varianceW}, we have
\begin{align*}
    \langle D_{\text{KL}}(\rho[h, t_i] || \rho[h+h, t_i])\rangle_{w} &= -\int\mathcal{D}h\{\frac{1}{\rho}\int d^3x\frac{\delta\rho}{\delta h_{ij}}\langle w_{ij}(x)\rangle+\int d^3xd^3y\frac{1}{2\rho}\frac{\delta\rho}{\delta h_{ij}}\langle w_{ij}(x)w_{kl}(y)\rangle\frac{\delta\rho}{\delta h_{kl}} \}\\
    &= \frac{\hbar\Delta t}{2} \int\mathcal{D}h d^3x\{\frac{16\pi GN}{\rho}\frac{\delta\rho}{\delta h_{ij}}\frac{G_{ijkl}}{\sqrt{h}}\frac{\delta\rho}{\delta h_{kl}}\}.
\end{align*}
Summing over the contribution of the relative entropy of all hypersurfaces $\Sigma_i$, and taking the limit $\Delta t\to 0$, we arrive at \eqref{FisherInfo2}. If the information metrics are defined using the Tsallis divergence, that is, $I_f^{\alpha}$ as defined in (\ref{TDivergence}), we have a slightly different Taylor expansion.
\begin{align*}
    \int \mathcal{D}h \frac{\rho^\alpha[h, t_i]}{\rho^{\alpha-1} [h+w,t_i]} 
    =& \int\mathcal{D}h \{\rho+(1-\alpha)[\int d^3x\frac{\delta\rho}{\delta h_{ij}}w_{ij}(x) ] +\frac{1}{2}\alpha(\alpha-1)(\frac{1}{\rho}[\int d^3x\frac{\delta\rho}{\delta h_{ij}}w_{ij}(x)]^2 \}.
\end{align*}
Substitute the above expansion into (\ref{TDivergence}), and take the expectation values $\langle\cdot\rangle_{w}$. Again, use the identities in \eqref{varianceW}, $I_f^{\alpha}$ is simplified as
\begin{align}
\label{RDivergence2}
    I_f^{\alpha} 
    &=\sum_{i=0}^{N-1}\langle\frac{1}{\alpha-1}(\int \mathcal{D}h\frac{\rho^\alpha[h, t_i]}{\rho^{\alpha-1} [h+w,t_i]} - 1)\rangle_w = \sum_{i=0}^{N-1}\frac{\alpha}{2} \int \mathcal{D}h\int d^3xd^3y\frac{1}{\rho}\frac{\delta\rho}{\delta h_{ij}}\langle w_{ij}(x)w_{kl}(y)\rangle\frac{\delta\rho}{\delta h_{kl}} \\
    &= \alpha\frac{\hbar\Delta t}{2} \int\mathcal{D}h d^3x\{\frac{16\pi GN}{\rho}\frac{\delta\rho}{\delta h_{ij}}\frac{G_{ijkl}}{\sqrt{h}}\frac{\delta\rho}{\delta h_{kl}}\} = \alpha I_f.
\end{align}

\section{Derivation of the Wheeler-DeWitt Equation}
\label{appendix:SE}
To derive equation \eqref{var7}, we perform the variation procedure on $A_t$ with respect to $S$. Only terms containing $S$ are involved in the variation. Therefore, 
\begin{equation}
  \delta_sA_t =  \int dt\mathcal{D}h\{\rho[\frac{\partial \delta'S}{\partial t} +\lambda^{(1)}\frac{\partial \delta'S}{\partial t}+\int d^3x \frac{32\pi GN}{\sqrt{h}}\frac{\delta (\delta'S)}{\delta h_{ij}}G_{ijkl}\frac{\delta S}{\delta h_{kl}} +\int d^3x (-2N_i+\lambda_i^{(2)})D_j(\frac{\delta (\delta'S)}{\delta h_{ij}})]\}.
\end{equation}
Note that the symbol $\delta'$ refers to the variation over the functional $S$ while $\delta$ refers to the variation over the field variable $h$. By choosing the Lagrangian multiplier $\lambda_i^{(2)}=2N_i$, the last term vanishes. Integration by part leads to the following result,
\begin{equation}
    \delta_sA_t =  -\int dt\mathcal{D}h\{(1 +\lambda^{(1)})\frac{\partial \rho}{\partial t}+32\pi G\int d^3x N\frac{\delta}{\delta h_{ij}}(\rho\frac{G_{ijkl}}{\sqrt{h}}\frac{\delta S}{\delta h_{kl}})\}\delta'S
\end{equation}
By demanding $\delta'S_t = 0$ for arbitrary $\delta'S$, we obtain \eqref{var7}. Next, we take the variation of $A_t$ with respect to $\rho$.
\begin{equation}
\label{F5}
\begin{split}
  \delta_\rho A_t&=  \int dt\mathcal{D}h\{[(1+\lambda^{(1)})\frac{\partial S}{\partial t}+\int d^3x N\mathcal{H}_\perp +\int d^3x (-2N_i+\lambda_i^{(2)})D_j(\frac{\delta S}{\delta h_{ij}})]\delta'\rho\\
  &+\delta_\rho (\rho F[\rho])+\lambda^{(3)}(\frac{\partial \rho}{\partial t}\delta'\rho + \rho\frac{\partial \delta'\rho}{\partial t})+[\delta'\rho\int d^3x\lambda_i^{(4)}D_j(\frac{\delta \rho}{\delta h_{ij}})+\rho\int d^3x\lambda_i^{(4)}D_j(\frac{\delta (\delta'\rho)}{\delta h_{ij}})]\}.
\end{split}
\end{equation}
Since we have chosen $\lambda_i^{(2)}=2N_i$, the third term vanishes. Integration by part shows that the fifth term also vanishes. The fourth term is equivalent to the variation of $I_f$ given in (\ref{FisherInfo}) with a small arbitrary change of  $\rho$, $\delta'\rho$,
\begin{align}
    \delta'_\rho (\frac{\hbar}{2}I_f) =& \frac{\hbar^2}{4}\int dt\mathcal{D}hd^3x\frac{16\pi GN}{\sqrt{h}}\{-\frac{\delta'\rho}{\rho^2}\frac{\delta \rho}{\delta h_{ij}}G_{ijkl}\frac{\delta \rho}{\delta h_{kl}}+\frac{1}{\rho}\frac{\delta (\delta'\rho)}{\delta h_{ij}}G_{ijkl}\frac{\delta \rho}{\delta h_{kl}} + \frac{1}{\rho}\frac{\delta \rho}{\delta h_{ij}}G_{ijkl}\frac{\delta (\delta'\rho)}{\delta h_{kl}}\}\\
    \label{E7}
    &=\frac{\hbar^2}{4}\int dt\mathcal{D}hd^3x(16\pi GN)\{\frac{1}{\rho^2}\frac{\delta \rho}{\delta h_{ij}}\frac{G_{ijkl}}{\sqrt{h}}\frac{\delta \rho}{\delta h_{kl}}-\frac{2}{\rho}\frac{\delta }{\delta h_{ij}}\frac{G_{ijkl}}{\sqrt{h}}\frac{\delta \rho}{\delta h_{kl}}\}\delta'\rho.
\end{align}
Defining $R=\sqrt{\rho}$, one can verify that
\begin{equation}
\label{BohmP}
    -\frac{4}{R}\frac{\delta }{\delta h_{ij}}\frac{G_{ijkl}}{\sqrt{h}}\frac{\delta R}{\delta h_{kl}}=\frac{1}{\rho^2}\frac{\delta \rho}{\delta h_{ij}}\frac{G_{ijkl}}{\sqrt{h}}\frac{\delta \rho}{\delta h_{kl}}-\frac{2}{\rho}\frac{\delta }{\delta h_{ij}}\frac{G_{ijkl}}{\sqrt{h}}\frac{\delta \rho}{\delta h_{kl}}. 
\end{equation}
Inserting it into \eqref{E7} gives
\begin{equation}
    \delta'_\rho (\frac{\hbar}{2}I_f) = -16\pi G\hbar^2\int dt\mathcal{D}hd^3x \frac{N}{R}\frac{\delta }{\delta h_{ij}}(\frac{G_{ijkl}}{\sqrt{h}}\frac{\delta R}{\delta h_{kl}})\delta'\rho.
\end{equation}
The last term in \eqref{F5} is more difficult to calculate. The second expression is evaluated as follows:
\begin{equation}
\label{F9}
    \int\mathcal{D}h d^3x[\rho\lambda_i^{(4)}D_j(\frac{\delta (\delta'\rho)}{\delta h_{ij}})]= \int\mathcal{D}h d^3x\{\frac{\delta}{\delta h_{ij}}[\rho D_j(\lambda_i^{(4)})]\}\delta'\rho =\int\mathcal{D}h d^3x\{\frac{\delta \rho}{\delta h_{ij}}[ D_j(\lambda_i^{(4)})] + \rho\frac{\delta }{\delta h_{ij}}[ D_j(\lambda_i^{(4)})]\}\delta'\rho
\end{equation}
The first term in \eqref{F9} and the first expression in the last term of \eqref{F5} cancel out after integration by part. Thus, the last term of \eqref{F5} becomes the second term in \eqref{F9}. Choosing the Lagrangian multiplier $\lambda_i^{(4)}=\sqrt{h}M_i$, it becomes
\begin{equation}
\label{F10}
    \int\mathcal{D}h\{\rho\int d^3x\frac{\delta }{\delta h_{ij}}[ D_j(\sqrt{h}M_i)]\}\delta'\rho.
\end{equation}
It can be shown that $\frac{\delta }{\delta h_{ij}}[ D_j(\sqrt{h}M_i)]=2\sqrt{h}D_iM^i$. But   $\sqrt{h}D_iM^i=\partial_i(\sqrt{h}M^i)$ is a total derivative. Thus, \eqref{F10}, and consequently the last term in \eqref{F5}, is a boundary term and can be ignored. Putting everything together, we find that
\begin{equation}
      \delta_\rho A_t=  \int dt\mathcal{D}h\{(1+\lambda^{(1)})\frac{\partial S}{\partial t}+\int d^3x [N\mathcal{H}_\perp -16\pi G\hbar^2 \frac{N}{R}\frac{\delta }{\delta h_{ij}}(\frac{G_{ijkl}}{\sqrt{h}}\frac{\delta R}{\delta h_{kl}})]\}\delta'\rho.
\end{equation}
Requiring $\delta_\rho A_t=0$ for the arbitrary $\delta'\rho$ leads to \eqref{var8}.

To combine \eqref{var11} and \eqref{var12} into the Wheeler-DeWitt equation, we define $\Psi[h, t]=\sqrt{\rho[h, t]}e^{iS}$, and evaluate the second order functional derivatives with respect to the field variable $h$.
\begin{align*}
    \frac{\delta \Psi}{\delta h_{ij}} =& \frac{1}{2\rho}\frac{\delta \rho}{\delta h_{ij}}\Psi +i\frac{\delta S}{\delta h_{ij}}\Psi,\\
    \frac{\delta }{\delta h_{ij}}\frac{G_{ijkl}}{\sqrt{h}}\frac{\delta}{\delta h_{kl}} \Psi&= \{\frac{i}{\hbar}(\frac{1}{\rho}\frac{\delta }{\delta h_{ij}}(\rho\frac{G_{ijkl}}{\sqrt{h}}\frac{\delta S}{\delta h_{kl}}) )-(\frac{1}{\hbar^2}\frac{\delta S}{\delta h_{ij}}\frac{G_{ijkl}}{\sqrt{h}}\frac{\delta S}{\delta h_{kl}}+\frac{1}{4\rho^2}\frac{\delta \rho}{\delta h_{ij}}\frac{G_{ijkl}}{\sqrt{h}}\frac{\delta\rho}{\delta h_{kl}}-\frac{1}{2\rho}\frac{\delta }{\delta h_{ij}}\frac{G_{ijkl}}{\sqrt{h}}\frac{\delta\rho}{\delta h_{kl}})\}\Psi\\
    &=\{\frac{i}{\hbar}(\frac{1}{\rho}\frac{\delta }{\delta h_{ij}}(\rho\frac{G_{ijkl}}{\sqrt{h}}\frac{\delta S}{\delta h_{kl}}) )-(\frac{1}{\hbar^2}\frac{\delta S}{\delta h_{ij}}\frac{G_{ijkl}}{\sqrt{h}}\frac{\delta S}{\delta h_{kl}}-\frac{1}{R}\frac{\delta }{\delta h_{ij}}(\frac{G_{ijkl}}{\sqrt{h}}\frac{\delta R}{\delta h_{kl}}))\}. 
\end{align*}
Substituting \eqref{var11} and \eqref{var12} into the above equation results in \eqref{Psi3}.

\end{document}